\documentclass[runningheads]{llncs}
\usepackage[T1]{fontenc}
\usepackage{graphicx}
\usepackage{url} 

\usepackage{float}
\usepackage{tabularx}   
\usepackage{array}      
\usepackage{amsmath}
\usepackage[shortcuts]{extdash} 
\usepackage{svg}

\usepackage[most]{tcolorbox}
\tcbuselibrary{breakable}

\definecolor{myblue}{HTML}{4B96F7}

\begin{document}
\title{
From "Help" to Helpful: A Hierarchical Assessment of LLMs in Mental e-Health Applications
}
\titlerunning{
Hierarchical LLM Assessment in e-Mental Health
}
%
\author{%
  Philipp Steigerwald\inst{1}\orcidID{0009-0002-5564-4279} \\
  \and
  Jens Albrecht\inst{1}\orcidID{0000-0003-4070-1787}
}

\authorrunning{P.\,Steigerwald and J.\,Albrecht}

%
\institute{%
  Technische Hochschule Nürnberg – Georg Simon Ohm, Nürnberg, Germany\\
  \email{\{philipp.steigerwald,jens.albrecht\}@th-nuernberg.de}\\
}

\hyphenation{
  Sau-er-kraut-LM
  as-ses-sors
  coun-sel-ling
  do-main
  train-ing
}

\maketitle              
\begin{abstract}
Psychosocial online counselling frequently encounters generic subject lines that impede efficient case prioritisation.
This study evaluates eleven large language models generating six-word subject lines for German counselling emails through hierarchical assessment\---first categorising outputs, then ranking within categories to enable manageable evaluation.
Nine assessors (counselling professionals and AI systems) enable analysis via Krippendorff's $\alpha$, Spearman's $\rho$, Pearson's $r$ and Kendall's $\tau$.
Results reveal performance trade-offs between proprietary services and privacy-preserving open-source alternatives, with German fine-tuning consistently improving performance.
The study addresses critical ethical considerations for mental health AI deployment including privacy, bias and accountability.

\end{abstract}

\section{Introduction}
Large Language Models (LLMs) have emerged as powerful tools for supporting mental health services \cite{guo_large_2024}, particularly in psychosocial online counselling where they could ease counsellor workload \cite{steigerwald_comparing_2025}.
This first-line intervention offers accessible support to individuals experiencing psychosocial challenges \cite{lattie_overview_2022,aspvall_effect_2021,ierardi_effectiveness_2022}.
However, asynchronous email counselling faces a persistent challenge: clients typically use generic subject lines like "Help" or "Problem" that fail to convey the nature or urgency of their concerns, impeding efficient triage when counsellors manage multiple cases \cite{steigerwald_comparing_2025}.

Building on the initial investigation with six assessors \cite{steigerwald_comparing_2025}, this expanded study confirms the original findings with enhanced methodological rigour through three additional AI assessors, introduces ranking-based analysis alongside categorical assessment and provides comprehensive ethical considerations for deployment in sensitive counselling contexts.

Text summarisation represents a foundational capability of contemporary language models \cite{zhang_systematic_2024}.
Well-crafted subject lines function as ultra-short summaries, distilling email content to orient recipients immediately.
Condensing counselling emails into meaningful subject lines exemplifies a summarisation challenge where LLMs have demonstrated proficiency across domains \cite{zhang_benchmarking_2024}, yet the psychosocial context introduces unique complexities: models must navigate sensitive emotional content whilst respecting client dignity.

Data protection requirements fundamentally shape technology choices in this domain.
Client communications contain highly sensitive information demanding exceptional safeguards.
Locally deployed open-source LLMs offer a promising path to process data within institutional boundaries, minimising exposure to external processors, whilst proprietary systems require scrutiny regarding privacy and regulatory alignment.
Ethical considerations\---encompassing bias mitigation, transparency and preservation of professional judgement\---motivate systematic evaluation of open-source alternatives alongside proprietary baselines.

\begin{figure}[htb]
    \centering
    \includegraphics[width=1\linewidth]{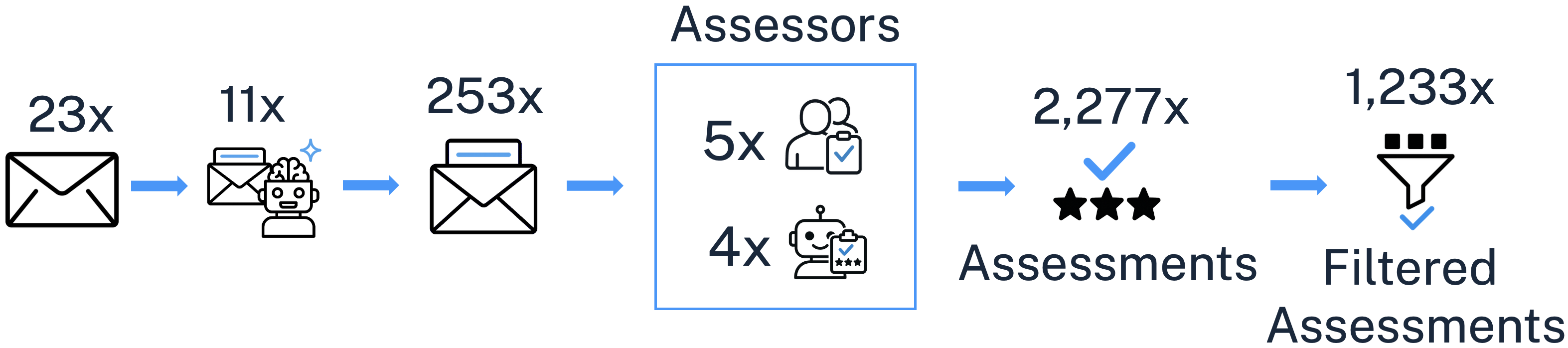}
    \caption{Evaluation workflow showing the progression from 23 counselling email threads through LLM processing (11 models generating 253 subject lines), assessment by nine evaluators (five human counselling professionals and four AI systems producing 2,277 assessments), to final filtered assessments (1,233) after applying inter-rater reliability thresholds.}
    \label{fig:workflow}
\end{figure}

This study investigates automated subject line generation through a hierarchical assessment framework.
Eleven LLMs\---spanning proprietary and open-source families, including German-tuned and quantised variants\---generated subject lines for 23 counselling email threads.
Nine assessors (five counselling professionals and four AI systems) assessed each output using a two-step process: first categorising into quality tiers (Good, Fair, Poor), then ranking within categories, reducing cognitive load whilst enabling fine-grained discrimination between models.
Figure~\ref{fig:workflow} illustrates the complete evaluation workflow from email processing through filtered assessments.

The investigation addresses five research questions:
\begin{enumerate}
    \item Can LLMs effectively condense counselling emails into concise, meaningful subject lines?
    \item Does language-specific fine-tuning (German) improve output quality across both evaluation methods?
    \item How do model size and quantisation affect subject line quality?
    \item Do human or AI assessors demonstrate higher inter-rater agreement and what does this reveal about evaluation consistency?
    \item To what extent does ranking analysis confirm categorical hierarchies whilst revealing subtle performance differences?
\end{enumerate}

This work contributes: 
(1) an expanded follow-up to the earlier study \cite{steigerwald_comparing_2025}, broadening systematic evaluation of LLM-generated subject lines for psychosocial counselling; 
(2) validation of language-specific adaptation value in mental-health applications; 
(3) a hierarchical evaluation framework leveraging complementary cognitive strengths of human experts and LLMs; 
(4) identification of privacy-friendly, open-source model alternatives achieving competitive quality; and 
(5) methodological foundations for assessing AI tools when performance and ethical considerations jointly determine suitability.

\section{Related Work}  
The automated generation of subject lines for psychosocial counselling emails intersects text summarisation, mental health applications and evaluation methodologies, establishing the theoretical foundation for the hierarchical assessment approach.

\subsection{Text Summarisation and Ultra-Short Summary Generation}
Text summarisation has evolved from extractive to abstractive approaches capturing essential meaning in novel formulations \cite{zhang_systematic_2024}.
Whilst traditional summarisation targets paragraph-length outputs, recent work pushes toward ultra-short summaries distilling complex information into minimal text.
Email summarisation presents unique challenges, with automated subject line generation pioneered through extracting salient sentences then rewriting them concisely \cite{zhang_this_2019}.
This demonstrated that effective subject lines require contextual understanding beyond brevity\---particularly challenging in psychosocial counselling where emotional nuance must be preserved.

The healthcare domain has shown particular interest in ultra-short summarisation.
Physicians benefit from automatically generated ultra-short abstracts when screening scientific literature, reducing cognitive load whilst maintaining decision quality \cite{kocbek_generating_2022}\---directly motivating this approach for counsellors managing multiple clients.
Recent advances handle long documents through hierarchical summarisation by segmenting, compressing and re-synthesising content \cite{yin_novel_2024}.
This principle of hierarchical processing informs the evaluation approach\---first categorising outputs broadly, then ranking within categories for finer discrimination.

\subsection{LLMs in Mental Health Applications}
LLM applications in mental health contexts fall into two categories: direct client interaction and counsellor support tools.
Direct interaction systems including ChatCounselor \cite{liu_chatcounselor_2023}, Psy-LLM \cite{lai_supporting_2024} and MentalBlend \cite{gu_mentalblend_2024} attempt to simulate counselling conversations but face fundamental limitations.
Current LLMs lack the emotional intelligence, ethical grounding and contextual understanding necessary for autonomous counselling practice \cite{chiu_computational_2024,koutsouleris_promise_2022}.

More promising are augmentation tools supporting human counsellors through response suggestions, pattern identification, or information organisation \cite{fu_enhancing_2023,hsu_helping_2023}.
Subject line generation fits this augmentation paradigm, enhancing efficiency without replacing clinical judgement and within this paradigm CAIA condenses information from long email threads to support triage and documentation \cite{steigerwald_enhancing_2024}. 
The mental health context imposes unique evaluation requirements, as even technically correct outputs can be inappropriate \cite{vowels_ai_2024}, necessitating domain-expert evaluation of both linguistic quality and counselling appropriateness.

\subsection{Evaluation Methodologies for LLM Outputs}
Traditional metrics like BLEU \cite{papineni_bleu_2002}, ROUGE \cite{lin_rouge_2004} and BERTScore \cite{zhang_bertscore_2019} often fail to capture mental health language nuances, making human evaluation the gold standard for sensitive domains \cite{tam_framework_2024}.

Healthcare studies demonstrate successful evaluation approaches, achieving high inter-rater reliability among medical professionals evaluating LLM diagnostic capabilities \cite{khan_comparison_2024}.
Krippendorff's $\alpha$ has assessed AI-generated counselling responses with robust results \cite{rudolph_ai-based_2024}.
Research exploring LLMs as assessors shows promise in educational assessment \cite{hackl_is_2023} and argument quality analysis \cite{mirzakhmedova_are_2024}, though domain-expert oversight remains essential for safety-critical applications \cite{tam_framework_2024}.

Recent methodological advances support combining rating and ranking approaches.
The "ceiling effect" problem occurs when competent LLMs cluster at rating scale tops, preventing meaningful discrimination \cite{thomson_mostly_2024}.
Ranking-based evaluation leverages human cognitive strengths in comparative judgement \cite{gao_human-like_2023}, with structured pairwise comparisons improving agreement between automated and expert evaluations for summarisation \cite{liu_g-eval_2023}.
Mixing categorical ratings with rankings yields more robust evaluations than either method alone \cite{belz_comparing_2010,liusie_llm_2024}.
Healthcare contexts particularly benefit from hybrid approaches where categorical ratings enforce safety thresholds before rankings differentiate among acceptable options \cite{tam_framework_2024,balloccu_ask_2024}.

Building on our initial investigation \cite{steigerwald_comparing_2025}, which demonstrated the feasibility of LLM-generated subject lines for psychosocial counselling and the benefits of German-language fine-tuning with five human assessors and one AI system, this expanded study strengthens and extends these findings.
The current work confirms the original results with an expanded assessor panel of nine (five counselling professionals and four AI systems), providing enhanced statistical reliability.
Additionally, this study introduces ranking-based analysis alongside categorical assessment, leveraging the cognitive advantages of categorisation-before-ranking to enable finer discrimination between models with similar performance.
The expanded use of AI assessors enables systematic comparison of human versus AI evaluation consistency in mental health contexts.
Furthermore, this work provides comprehensive ethical analysis of deploying such tools in sensitive counselling environments, examining privacy trade-offs between open-source and proprietary systems, bias implications and accountability frameworks essential for responsible implementation.
The following methodology extends the hierarchical assessment framework, providing both methodological validation and practical deployment guidance for sensitive AI applications.

\section{Methodology} 
\label{sec: Methodology}
This study evaluates eleven LLMs generating subject lines for 23 German psychosocial counselling email threads, producing 253 outputs.
Nine assessors\---five counselling professionals and four AI systems\---assessed each subject line using a hierarchical approach that first categorises outputs then ranks within categories.
The resulting 2,277 assessments provide both absolute quality measures and nuanced performance distinctions, capturing how models perform independently and in relation to one another.

\subsection{Study Design}
\label{sec: Study Design}
The evaluation framework employs a hierarchical assessment approach that aligns with natural human decision-making processes.
Rather than requiring assessors to directly rank all eleven subject lines for each email thread\---a cognitively demanding task prone to inconsistency\---the framework divides assessment into two manageable stages.
Additionally, assessing subject lines in direct comparison proves easier than categorising individual outputs in isolation, as relative quality differences become more apparent when alternatives are visible.

\begin{figure}[htb]
    \centering
    \includegraphics[width=1\linewidth]{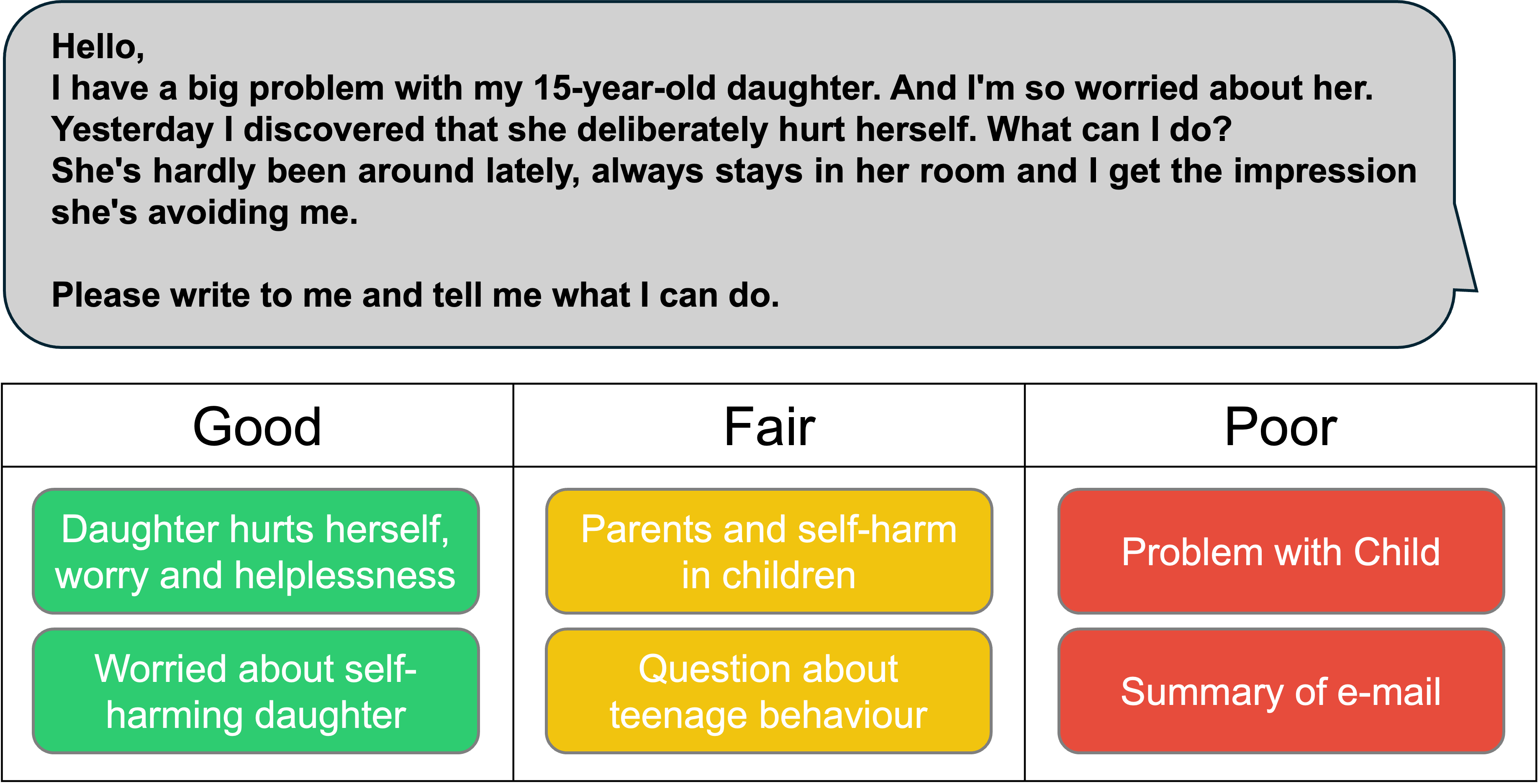}
    \caption{Hierarchical categorisation example for a counselling email about self-harm concerns, demonstrating quality distinctions between Good (specific), Fair (partial) and Poor (generic) subject lines (adapted from \cite{steigerwald_comparing_2025}).}
    \label{fig:hierarchical_evaluation_example}
\end{figure}

In the first stage, assessors categorise each subject line into one of three quality tiers: Good, Fair, or Poor.
As illustrated in Figure \ref{fig:hierarchical_evaluation_example}, Good ratings indicate subject lines that accurately capture the specific concern (self-harm) and emotional context (worry, helplessness).
Fair ratings suggest partial success, identifying relevant themes but lacking precision\---for instance, mentioning "teenage behaviour" without specifying the critical self-harm element.
Poor ratings identify outputs that remain generic or merely restate that an email exists without providing actionable information.

Following categorisation, assessors rank all subject lines within each quality tier from best to worst.
Assessing subject lines in direct comparison proves easier than rating individual outputs in isolation, as relative quality differences become immediately apparent when alternatives are visible.
However, directly ordering eleven subject lines simultaneously proved cognitively overwhelming for assessors.
Therefore, the hierarchical approach was adopted: assessors first sort outputs into Good, Fair and Poor categories, then rank only within these smaller groups.
This pre-sorting transforms an overwhelming eleven-way comparison into manageable subsets of typically three to four items, making the ranking task feasible whilst preserving the advantages of comparative assessment.

\subsection{Email Thread Construction}
The evaluation dataset comprised 23 German-language email threads specifically created for this study.
These threads were constructed by experienced psychosocial counselling professionals who drew upon their extensive practice knowledge to create realistic scenarios.
Each thread represents a typical counselling situation that could plausibly occur in practice, encompassing common client concerns such as relationship difficulties, family conflicts, mental health challenges and life transitions.
This synthetic approach was necessitated by stringent data protection requirements governing authentic counselling communications.
Whilst this methodology produces what might be termed a "silver standard" rather than authentic client emails, it ensures complete protection of client confidentiality whilst maintaining ecological validity through the practitioners' expertise in crafting representative scenarios.
The limited number of 23 threads reflects the substantial effort required for experienced counsellors to construct realistic, contextually rich email exchanges that accurately reflect the complexity of genuine counselling interactions.
Future research should prioritise obtaining genuinely anonymised or pseudonymised counselling communications to enhance ecological validity.
However, accessing such data remains exceptionally challenging due to the sensitive nature of mental health communications and corresponding regulatory frameworks.
The current approach represents a pragmatic balance between methodological rigour and ethical obligations.

\subsection{LLM Selection}
\label{sec: LLM Selection}

Model selection followed a systematic approach to examine performance across multiple dimensions: proprietary versus open-source architectures, standard versus German-adapted variants and different quantisation levels.
Table \ref{tab: models} presents the eleven models evaluated, spanning parameter counts from 8 billion to over 100 billion.

\begin{table}[ht]
  \caption{Overview of models evaluated in the study. Models marked as "Full" use full-precision weights, whilst Q4 and Q8 indicate 4-bit and 8-bit quantisation respectively.}
  \label{tab: models}
  \centering
\begin{tabularx}{\textwidth}{|>
                                {\raggedright\arraybackslash}
                               X
                              |l
                              |c|c|c|}

    \hline
    \textbf{Model Name} & \textbf{Abbreviation} & \textbf{Full} & \textbf{Q4} & \textbf{Q8} \\ \hline
    OpenAI GPT-3.5 Turbo      & GPT-3.5       & x &   &   \\ \hline
    OpenAI GPT-4o             & GPT-4o        & x &   &   \\ \hline
    Meta Llama 3.1 8b Instruct      & Meta-LM3.1-8b &   & x & x \\ \hline
    Vago SauerkrautLM Llama 3.1 8b Instruct& SKLM-LM3.1-8b &   & x & x \\ \hline
    Vago SauerkrautLM Llama 3 70b Instruct& SKLM-LM3-70b  &   & x &   \\ \hline
    Vago SauerkrautLM Mixtral 8×7b Instruct& SKLM-8x7b     &   & x & x \\ \hline
    Mistral Mixtral 8×7b Instruct      & Mistral-8x7b &   & x & x \\ \hline
  \end{tabularx}
\end{table}

OpenAI's GPT-3.5 Turbo and GPT-4o serve as proprietary benchmarks, representing state-of-the-art performance in text generation tasks \cite{shahriar_putting_2024,rao_single-document_2024}.
These cloud-based models provide the quality ceiling against which open-source alternatives must be measured.
The open-source selection includes Meta's Llama 3.1 8b \cite{dubey_llama_2024} and Mistral's Mixtral 8×7b \cite{jiang_mixtral_2024}, both widely adopted for their balance of performance and efficiency.
These base models enable direct comparison with their German-adapted counterparts from the SauerkrautLM family, developed by VAGO solutions to improve German-language understanding and generation.

SauerkrautLM variants span multiple architectures and sizes: Llama 3.1 8b, Llama 3 70b and Mixtral 8×7b.
This diversity enables analysis of how German adaptation benefits transfer across different model architectures and parameter counts.
The inclusion of both base and adapted versions allows direct measurement of language-specific fine-tuning impact.

Quantisation levels (4-bit and 8-bit) were selected to assess the feasibility of resource-constrained deployment.
Whilst full-precision models offer optimal quality, quantised variants require significantly less memory and computational power, potentially enabling local deployment on modest hardware.
The evaluation examines whether this efficiency gain comes at acceptable quality cost for counselling applications.

\subsection{Subject Line Generation}
\label{sec: Subject Generation}

The prompt engineering approach balanced specificity with flexibility to ensure consistent performance across diverse model architectures.
Each model received identical instructions structured to establish context, define the task precisely and specify output requirements.

The prompt design addressed several critical considerations.
First, it positioned the model as a counselling support tool rather than a counsellor, maintaining appropriate boundaries.
Second, it emphasised practical utility\---helping counsellors quickly identify and prioritise cases.
Third, it imposed strict constraints: maximum six words and JSON-formatted output, ensuring outputs remained concise and parseable.

\begin{tcolorbox}[colback=gray!5!white, colframe=myblue, title=Subject Line Generation Prompt, breakable, fontupper=\fontsize{8}{9}\selectfont]
\label{cb: prompt}
You are a specialised assistant for psychosocial online counselling. \\\\ 
Clients often approach counselling services with vague subject lines like "Help" or "Problem." Your role is to assist the counsellor by generating a precise and individual subject line for the client's first email. 
This helps the counsellor quickly grasp the main content of the request and respond efficiently, especially when managing multiple parallel cases. \\\\
Carefully read the client's first email and generate a concise subject line that clearly and understandably summarises the core issue of the request. The subject should be a maximum of 6 words and should not contain unnecessary formalities, enabling the counsellor to immediately gain a clear understanding of the issue.\\\\
The input consists of a complete email thread in chronological order. 
The email is formatted as: \{Role\} wrote on \{Date\} at \{Time\}: '{email Content}' \#\#\#.\\\\
The desired output is a JSON object containing one field: \{ "Subject": "Generated concise subject line" \}. \\\\
The subject line should concisely summarise the core content of the client's first message and avoid unnecessary formalities.
Do not use quotation marks or 'Subject:' in the generated subject.\\\\
Following the formatted email history is presented: \{\{complete\_email\_history\}\}\\End of email history.\\\\
Remember, you are a specialised assistant for psychosocial online counselling. Your task is to create concise and relevant subject lines that help the counsellor to quickly understand the client's issue.\\\\
Remember, your task is to read the client's first email in the thread and generate a short, concise subject line that accurately reflects the core content of the request.
\end{tcolorbox}

The prompt includes role reinforcement at beginning and end to maintain focus across longer email threads.
The structured format specification ensures consistent parsing across all model outputs, whilst the explicit examples of poor subject lines ("Help", "Problem") guide models away from generic responses.
All models utilised their native structured output capabilities to guarantee valid JSON formatting, eliminating post-processing requirements and ensuring fair comparison focused on content quality rather than format compliance.

\subsection{Assessor Configuration}
\label{sec: Assessor Configuration}
The evaluation employed nine assessors\---five human counselling professionals and four AI systems (o1-preview, o1, o3 and o4-mini).
This diverse panel enables analysis of both human expert judgment and AI evaluation consistency, whilst providing sufficient redundancy to identify and filter unreliable assessments.

The five human assessors, all experienced in German psychosocial online counselling, received standardised instructions emphasising practical utility over linguistic perfection.
Each assessor independently evaluated all 253 subject lines through the hierarchical framework: first categorising into Good, Fair, or Poor, then ranking within categories.
Assessment guidelines provided to human assessors specified:
\begin{quote}
\textit{Subject lines should be concise and individually tailored to the initial message. Each must summarise main content clearly in maximum 6 words, avoiding unnecessary formalities. Evaluation focuses on how precisely a subject line captures core content, enabling counsellors to quickly grasp the central concern.}
\end{quote}
To prevent bias, subject lines appeared in randomised order without model identifiers.
Assessors viewed each email's full context alongside all eleven generated subject lines, enabling both absolute quality judgments and relative comparisons within the same counselling scenario.

The four AI assessors followed an identical evaluation protocol implemented through structured prompting.
Each AI system received the email thread and all generated subject lines, with explicit instructions to categorise and rank according to the same criteria as human assessors.

\begin{tcolorbox}[colback=gray!5!white, colframe=myblue, title=AI Assessor Prompt (Excerpt), breakable, fontupper=\fontsize{8}{9}\selectfont]
\textbf{Role:} You act as a specialised model used in psychosocial online counselling to evaluate, categorise and rank automatically subject lines.

\textbf{Task:} Order the generated subject lines by quality and assign each suggestion to one of three buckets: `Good', `Fair', or `Poor'.

\textbf{Sequential Bucket Filling:}
\begin{itemize}
\item A subject line may only be placed in the `Good' bucket if no better subject line lies in the `Fair' or `Poor' buckets
\item Once a subject line enters `Fair', all subsequent subject lines must be `Fair' or `Poor' (the `Good' bucket closes)
\item Once a subject line enters `Poor', all subsequent subject lines must be `Poor' (the `Fair' bucket closes)
\item Empty buckets are possible if quality distribution requires it
\end{itemize}

\textbf{Output Format:} Valid JSON object with rankings:
\begin{verbatim}
{
  "rankings": [
    {
      "id": "...",
      "rank": n,
      "bucket": "Good/Fair/Poor"
    },
    ...
  ]
}
\end{verbatim}
\end{tcolorbox}

The prompt enforces sequential bucket filling to mirror human cognitive processes: once a subject line enters the Fair category, no subsequent items can be rated Good, regardless of quality.
This constraint ensures consistent application of quality thresholds across all assessments.
Each subject line received a unique hash identifier combining content, model origin and email thread ID.
This hash system enabled precise tracking whilst maintaining assessor blindness to model identity.

AI assessors processed each email thread through the following pipeline:
First, all subject lines for a given email were extracted and paired with their hash identifiers.
Second, the email conversation and subject line list were inserted into the prompt template.
Third, the AI model generated rankings with explanations, returning structured JSON output.
Fourth, validation checks ensured all subject lines were ranked exactly once and bucket sequencing remained valid.
The system implemented automatic retry logic for malformed responses, requesting up to 20 iterations until receiving valid JSON with complete rankings.
This rigorous validation prevented missing assessments or inconsistent bucket assignments, ensuring data integrity across all 2,277 individual evaluations.

Quality assurance measures included cross-validation of hash assignments, verification of complete coverage across all email threads and systematic checks for response format compliance.
The combination of human expertise and AI consistency provides a robust foundation for analysing both model performance and assessment methodology effectiveness.
\subsection{Statistical Analysis Framework}
\label{sec: Statistical Analysis}

This expanded study extends the statistical analysis from the original investigation \cite{steigerwald_comparing_2025} by incorporating additional metrics to evaluate both categorical ratings and complete rankings.
Whilst the initial work employed Krippendorff's alpha and Spearman correlation for categorical assessment, the current framework adds Kendall's tau for ranking consistency and Pearson correlation for method convergence analysis.
Four complementary metrics ensure robust evaluation across both assessment approaches, validating that observed patterns reflect genuine model performance rather than random variation.

Krippendorff's alpha serves as the foundational measure for inter-rater reliability when multiple assessors categorise subject lines into Good, Fair, or Poor ratings.
This metric accounts for chance agreement whilst simultaneously handling all nine assessors, providing a single reliability coefficient for the entire assessment panel:

\begin{equation}
\label{eq: krippendorff}
\alpha = 1 - \frac{D_o}{D_e}
\end{equation}

where $D_o$ represents observed disagreement between raters and $D_e$ represents expected disagreement by chance.
Unlike pairwise measures, Krippendorff's alpha evaluates agreement across all raters simultaneously, adjusting for the probability of chance concordance.
Values above 0.667 indicate acceptable reliability for drawing meaningful conclusions \cite{krippendorff_content_2018}.
The metric's suitability for ordinal data makes it particularly appropriate for categorical ratings where Good, Fair and Poor represent ordered quality levels.

Spearman's rank correlation coefficient evaluates monotonic relationships between assessors' categorical ratings by treating them as ordinal data.
Each categorical rating receives an ordinal value (Poor=1, Fair=2, Good=3), enabling correlation analysis between assessor pairs:

\begin{equation}
\label{eq: spearman}
\rho = 1 - \frac{6 \sum d_i^2}{n(n^2 - 1)}
\end{equation}

where $d_i$ represents rank differences between paired observations and $n$ denotes the number of observations.
This non-parametric measure captures whether assessors show similar patterns in their relative quality assessments, regardless of absolute rating differences.
Values range from -1 indicating perfect negative correlation to +1 representing perfect positive correlation.
High Spearman coefficients reveal that when one assessor rates a subject line higher than another, the second assessor tends to follow the same pattern.
Beyond pairwise agreement, Spearman correlation also quantifies the monotonic relationship between Good rating proportions and average ranks, providing initial validation of method convergence.

Kendall's tau specifically addresses ranking consistency when assessors order all eleven models from best to worst.
This metric quantifies ordinal association between complete rankings:

\begin{equation}
\label{eq: kendall}
\tau = \frac{2(C - D)}{n(n-1)}
\end{equation}

where $C$ represents concordant pairs (models ranked identically by both assessors), $D$ represents discordant pairs (models ranked oppositely) and $n$ equals the eleven models evaluated.
Kendall's tau proves particularly robust for ranking data as it considers all pairwise comparisons within the complete ordering.
Values range from -1 for perfect disagreement to +1 for perfect agreement, with positive values indicating consensus on relative model performance.
The metric's focus on pairwise concordance makes it ideal for evaluating whether assessors agree on which models perform better than others, even if exact rank positions differ slightly.

Pearson's correlation coefficient complements Spearman correlation to establish the nature of the relationship between categorical and ranking assessments.
This parametric measure quantifies the strength of linear association between Good rating proportions and average ranking positions:

\begin{equation}
\label{eq: pearson}
r = \frac{\sum_{i=1}^{n} (x_i - \bar{x})(y_i - \bar{y})}{\sqrt{\sum_{i=1}^{n} (x_i - \bar{x})^2 \sum_{i=1}^{n} (y_i - \bar{y})^2}}
\end{equation}

where $x_i$ and $y_i$ represent paired observations with means $\bar{x}$ and $\bar{y}$ respectively.
Whilst Spearman captures monotonic relationships regardless of linearity, Pearson specifically measures linear association strength.
By comparing Spearman and Pearson correlations for the same relationship, the analysis determines whether the association between Good ratings and average ranks follows a linear pattern.
Similar magnitudes between both coefficients indicate linearity, whilst divergence would suggest a non-linear monotonic relationship.
Strong negative correlation in both measures confirms that both assessment methods measure the same underlying quality construct through a linear relationship.
This dual-correlation approach proves essential for validating that the hierarchical assessment framework produces internally consistent, linearly related results.

Together, these four metrics establish comprehensive statistical rigour across multiple assessment dimensions.
Krippendorff's alpha ensures categorical reliability across all assessors, Spearman correlation reveals pairwise rating patterns and monotonic method relationships, Kendall's tau validates ranking consistency and Pearson correlation confirms linear method convergence.
The combination of Spearman and Pearson correlations specifically validates the linearity of the relationship between categorical and ranking assessments.
This multi-faceted approach distinguishes genuine performance differences from measurement noise, enabling confident conclusions about both model capabilities and assessment methodology effectiveness.
\section{Results}
\label{sec: Results}

The evaluation involved nine assessors\---five human counselling professionals and four AI systems (o1-preview, o1, o3 and o4-mini)\---who assessed 253 generated subject lines from eleven different LLMs across 23 email threads. 
This comprehensive assessment employed both categorical ratings (Good, Fair, Poor) and complete rankings (1-11), yielding 2,277 categorical ratings and 2,277 ranking positions for analysis.

The results are presented in four main sections. 
First, the data filtering process necessary to achieve acceptable inter-rater reliability is detailed, demonstrating how systematic filtering based on assessor agreement enhanced the reliability of subsequent analyses. 
Second, the categorical rating analysis examines model performance through traditional quality assessments and inter-rater agreement patterns. 
Third, the ranking-based analysis provides complementary insights through relative performance comparisons and assessor consistency measures. 
Finally, the correlation between both evaluation methods is explored, revealing how categorical and ranking assessments converge and diverge in their evaluation of model capabilities.

This multi-faceted analysis enables comprehensive understanding of LLM performance in generating subject lines for psychosocial counselling, whilst also examining systematic differences between human and AI assessors in applying both assessment methods. 
The inclusion of multiple AI assessors provides unique insights into the consistency of automated evaluation approaches compared to human expert judgment.

\subsection{Data Filtering and Reliability Enhancement}
\label{sec: data_filtering}

Initial analysis of the complete dataset revealed insufficient inter-rater reliability, with Krippendorff's $\alpha$ falling below the acceptable threshold of 0.667 required for drawing meaningful conclusions \cite{krippendorff_content_2018}.
In the original study \cite{steigerwald_comparing_2025}, applying the filtering procedure resulted in a Krippendorff's $\alpha$ of 0.685.
In the present expanded study, three additional ai assessors were included, increasing the total from six to nine. This change affected the agreement distribution and consequently, the filtering outcome, leading to a different optimisation curve and a revised agreement threshold compared to the original analysis.

\begin{figure}[htb] 
    \centering 
    \includegraphics[width=1\linewidth]{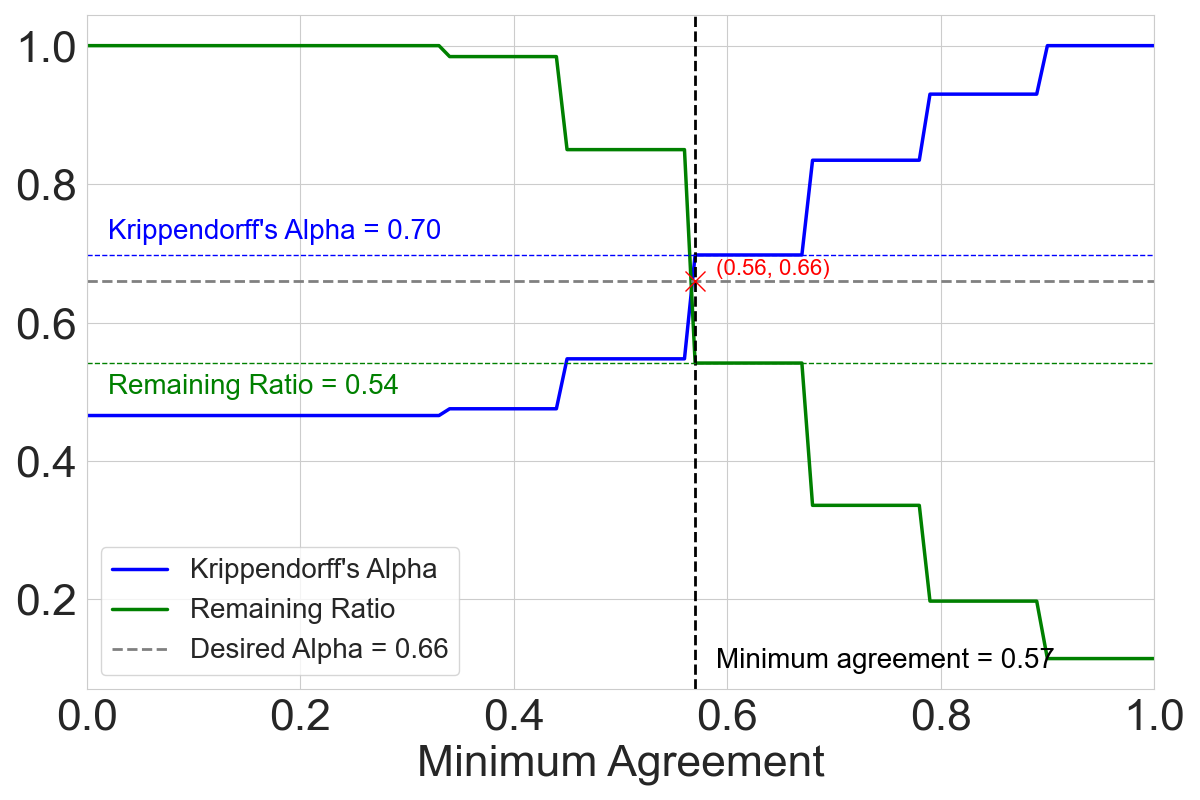} 
    \caption{Relationship between Krippendorff's $\alpha$ (blue), remaining data ratio (green) and minimum agreement threshold. The retained data proportion decreases as agreement thresholds rise, causing a corresponding increase in Krippendorff's $\alpha$ values. The red marker indicates where $\alpha$ exceeds 0.667 at 57\% agreement, achieving $\alpha$ = 0.70 whilst retaining 54.2\% of the data.} 
    \label{fig: set alpha} 
\end{figure}

Figure \ref{fig: set alpha} visualises this optimisation process, displaying the relationship between minimum agreement threshold, Krippendorff's $\alpha$ and data retention ratio.

The filtering algorithm incrementally evaluates agreement thresholds from 0 to 1. 
At each threshold level, the proportion of assessors agreeing on each subject line's categorisation is calculated. 
Subject lines meeting or exceeding the threshold are retained, whilst those with lower agreement are excluded. 
As the threshold increases, the dataset size decreases but overall reliability improves, as evidenced by the rising Krippendorff's $\alpha$ values.

With nine assessors\---expanded from six in the original study \cite{steigerwald_comparing_2025}\---agreement thresholds create discrete steps at multiples of approximately 11.11\%, explaining the stepwise changes in both metrics. 
The critical point occurs at an agreement threshold of 57\%, where Krippendorff's $\alpha$ reaches 0.70. 
At this threshold, subject lines require agreement from at least six of the nine assessors to remain in the dataset.

\begin{figure}[!ht]
    \centering    
    \includegraphics[width=1\linewidth]{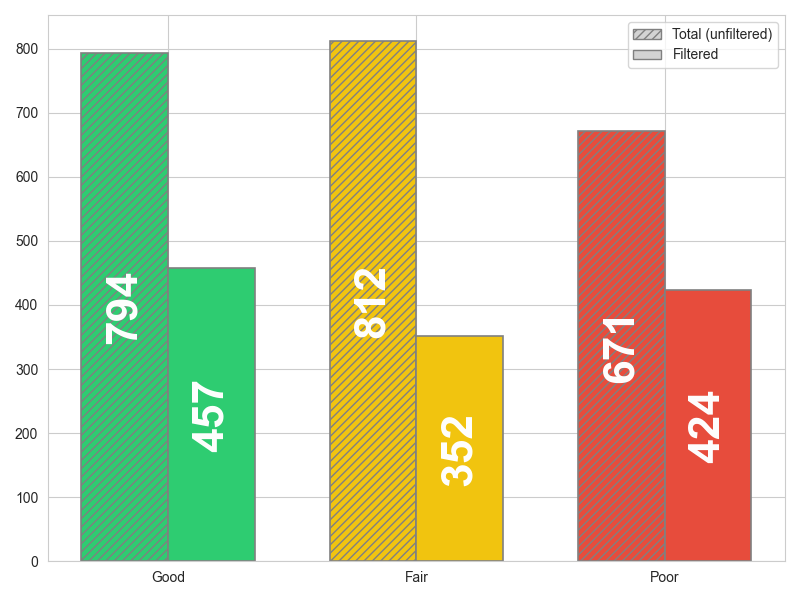}
    \caption{Distribution of ratings before (solid bars, n = 2,277) and after filtering (hatched bars, n = 1,233). Fair ratings showed the strongest reduction (56.7\%), followed by Good (42.4\%) and Poor ratings (36.8\%).}
    \label{fig: total bucket distribution}
\end{figure}

After filtering, 137 of the original 253 generated subject lines remained, resulting in a 45.8\% dataset reduction. 
Retention rates varied substantially across models: SauerkrautLM Mixtral 8×7b Q8 showed the highest retention at 73.9\% (17 of 23 subject lines), followed by Meta Llama 3.1 8b Q4 at 69.6\% (n=16) and GPT-4o at 65.2\% (n=15).
Mid-range retention occurred for Meta Llama 3.1 8b Q8 (60.9\%, n=14), Mixtral 8×7b Q4 (56.5\%, n=13) and SauerkrautLM Llama 3.1 8b Q4 (52.2\%, n=12). 
GPT-3.5 Turbo and SauerkrautLM Mixtral 8×7b Q4 both retained 47.8\% (n=11), whilst SauerkrautLM Llama 3 70b Q4 retained 43.5\% (n=10).
The lowest retention rates occurred for Mixtral 8×7b Q8 and SauerkrautLM Llama 3.1 8b Q8, both at 39.1\% (9 of 23).

The filtering process led to differential reductions across rating categories (Figure \ref{fig: total bucket distribution}). 
Fair ratings experienced the most substantial reduction of 56.7\% (from 812 to 352), whilst Poor ratings showed the highest retention at 63.2\% (from 671 to 424). 
Good ratings fell between these extremes with 57.6\% retention (from 794 to 457).
Post-filtering, inter-rater reliability reached Krippendorff’s $\alpha=0.70$, computed from 1,233 individual ratings and rankings (137 × 9 assessors) that formed the basis for subsequent analyses. The filtered dataset preserves the relative performance rankings of the evaluated models whilst reducing noise in the data.

\subsection{Categorical Rating Analysis}
\label{sec: rating_analysis}

\begin{figure}[htb]
    \centering
    \includegraphics[width=1.0\linewidth]{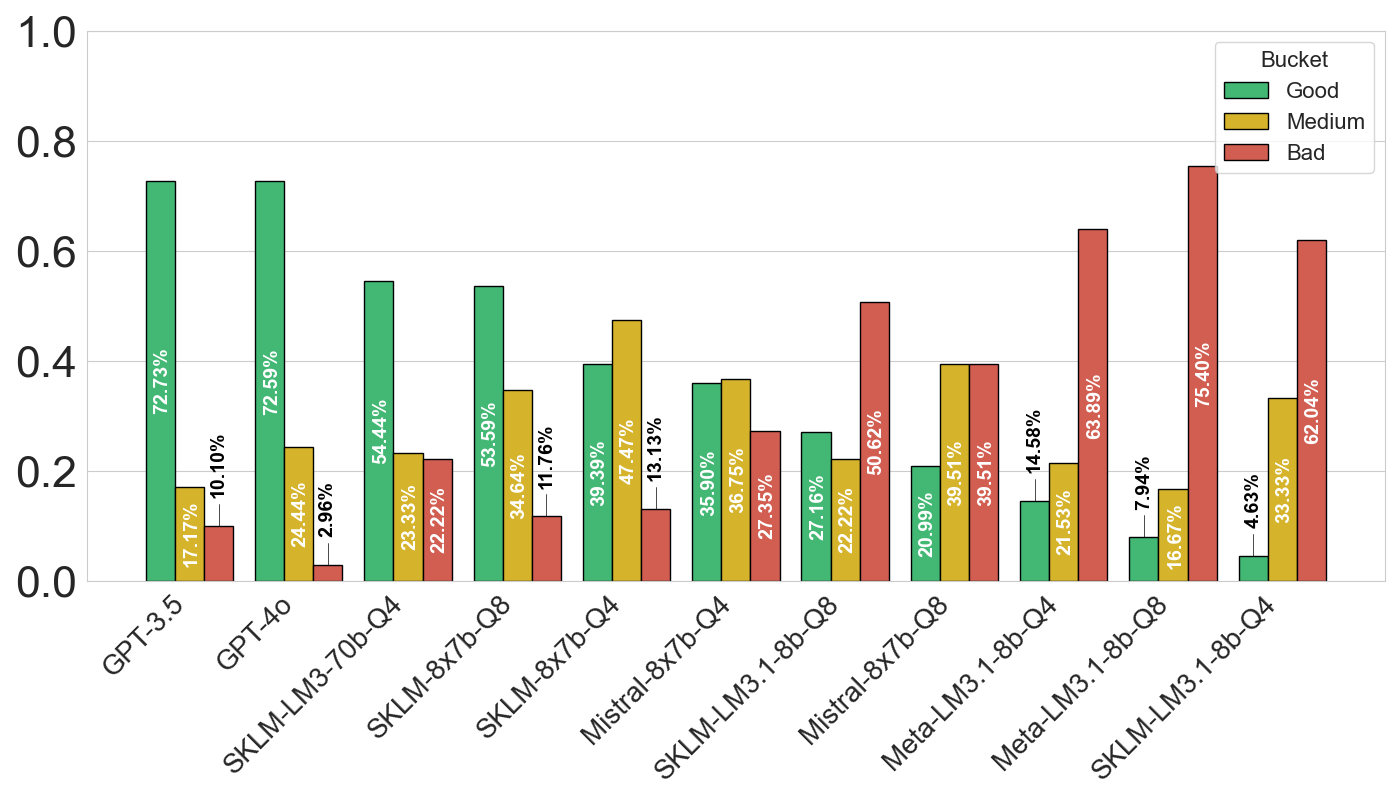}
    \caption{Distribution of filtered ratings (Good, Fair and Poor) across all evaluated models. Model names are abbreviated as follows: LM3.1 (Llama 3.1), LM3 (Llama 3) and 8×7b (Mixtral 8×7b). The suffixes Q4 and Q8 denote 4-bit and 8-bit quantisation, respectively.}
    \label{fig: weighted distribution}
\end{figure}

Following the filtering process, categorical performance was analysed through aggregate rating distributions and inter-rater agreement patterns.
Figure \ref{fig: weighted distribution} presents the distribution of Good, Fair and Poor ratings across all evaluated models.

\begin{figure}[htb]
    \centering
    \includegraphics[width=1\linewidth]{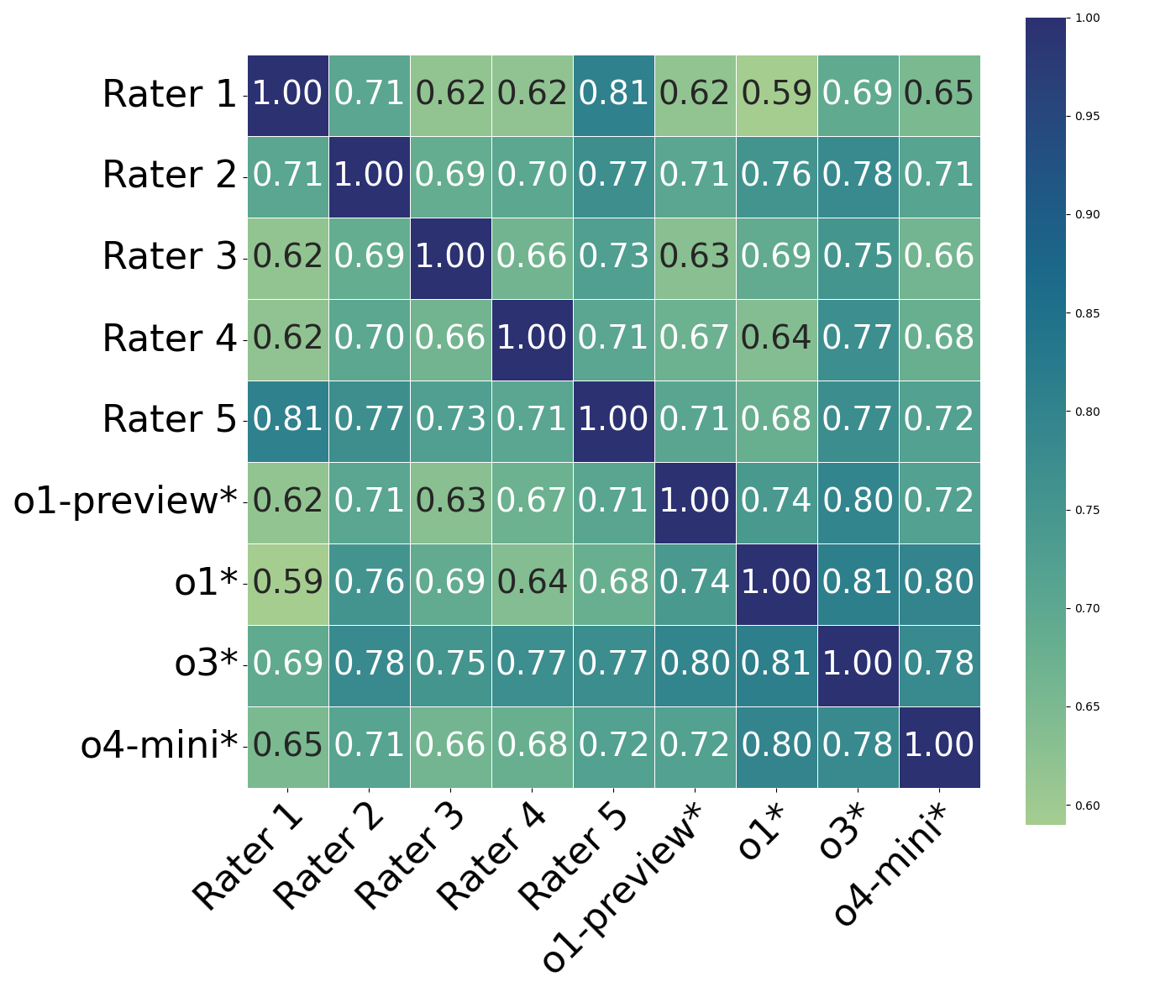}
    \caption{Pairwise Spearman correlation heatmap between the 9 assessors after filtering. Assessors 1-5 are human counselling professionals, whilst AI assessors (o1-preview, o1, o3 and o4-mini) are marked with asterisks.}
    \label{fig: filtered spearman}
\end{figure}

Both proprietary models, GPT-3.5 Turbo and GPT-4o, achieved 73\% Good ratings.
Among open-source models, SauerkrautLM Llama 3 70b Q4 and SauerkrautLM Mixtral 8×7b Q8 both achieved 54\% Good ratings.
SauerkrautLM Mixtral 8×7b Q8 showed 12\% Poor ratings compared to 22\% for SauerkrautLM Llama 3 70b Q4.
Mid-tier performance occurred for SauerkrautLM Mixtral 8×7b Q4 (39\% Good, 47\% Fair, 14\% Poor) and Mixtral 8×7b Q4 (36\% Good, 37\% Fair, 27\% Poor).
Lower performance was observed for SauerkrautLM Llama 3.1 8b Q8 (27\% Good) and Mixtral 8×7b Q8 (21\% Good).
The lowest performers were Meta Llama 3.1 8b Q4 (15\% Good, 64\% Poor), Meta Llama 3.1 8b Q8 (8\% Good, 75\% Poor) and SauerkrautLM Llama 3.1 8b Q4 (5\% Good, 62\% Poor).
Pairwise Spearman correlations between assessors after filtering are shown in Figure \ref{fig: filtered spearman}.

As shown in Figure~\ref{fig: filtered spearman}, human assessors (1-5) showed inter-rater correlations ranging from 0.62 to 0.81 (mean = 0.69). AI assessors demonstrated correlations between 0.74 and 0.81 (mean = 0.78). The highest AI-AI correlation occurred between o1 and o3 ($\rho = 0.81$). Human-AI correlations ranged from 0.59 to 0.78 (mean = 0.69). Rater 1 showed the lowest correlations with AI assessors (0.59-0.65), whilst Rater 2 demonstrated the highest (0.71-0.78).

\subsection{Ranking-Based Analysis}
\label{sec: ranking_analysis}

\begin{figure}[htb]
\centering
\includegraphics[width=1\linewidth]{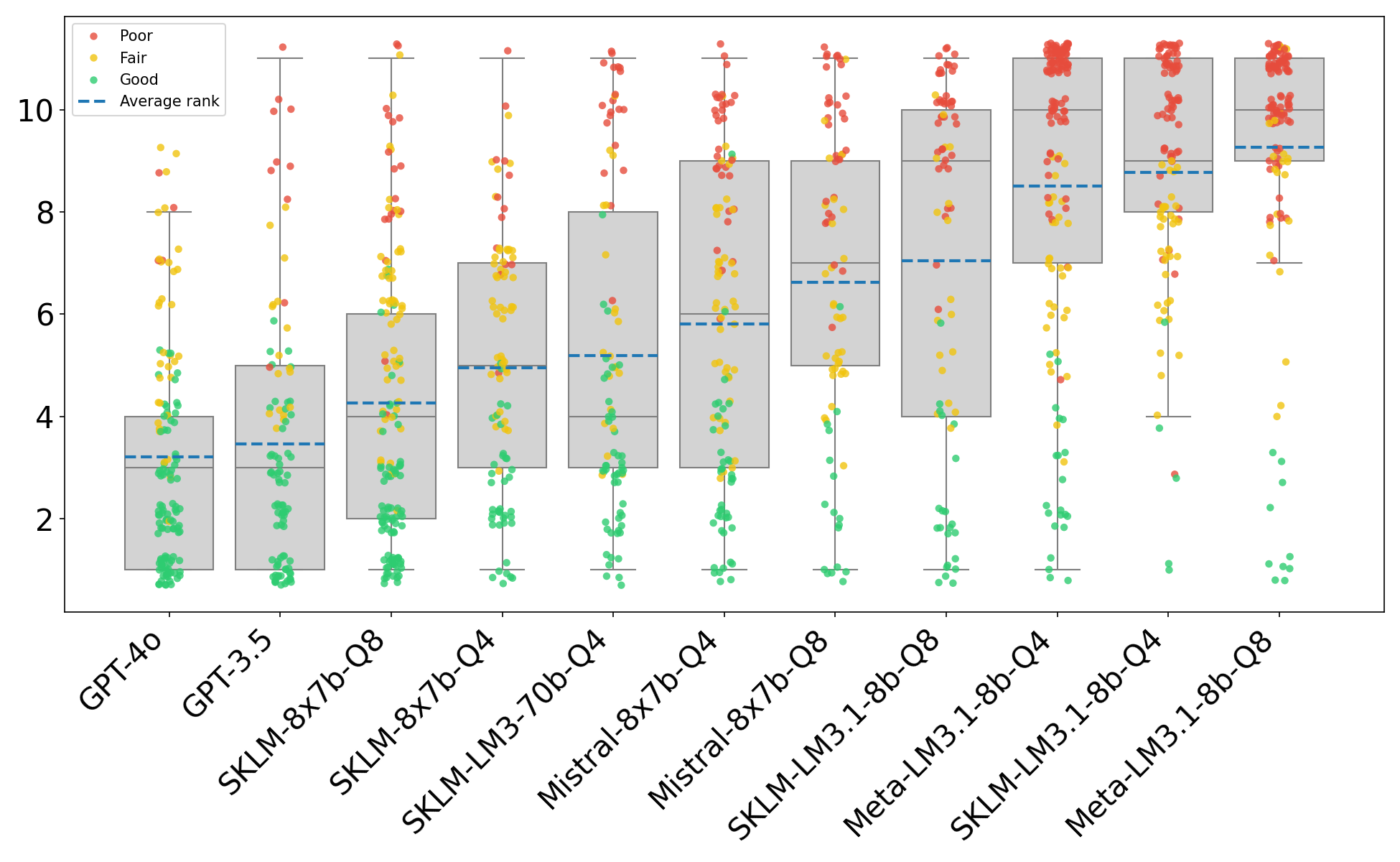}
\caption{Box plot showing the distribution of rankings for each LLM model across all assessors and filtered email threads.
Models are ordered by median rank performance, with lower ranks indicating better performance.
Coloured dots indicate categorical ratings (green = Good, yellow = Fair, red = Poor).}
\label{fig:rankings_boxplot}
\end{figure}

Each of the nine assessors produced complete rankings of the eleven LLMs for all 137 filtered subject lines, ordering them from rank 1 (best) to rank 11 (worst).
This yielded 1,233 ranking positions for analysis.
Figure \ref{fig:rankings_boxplot} presents the distribution of ranks achieved by each model.

\begin{figure}[htb]
\centering
\includegraphics[width=1\linewidth]{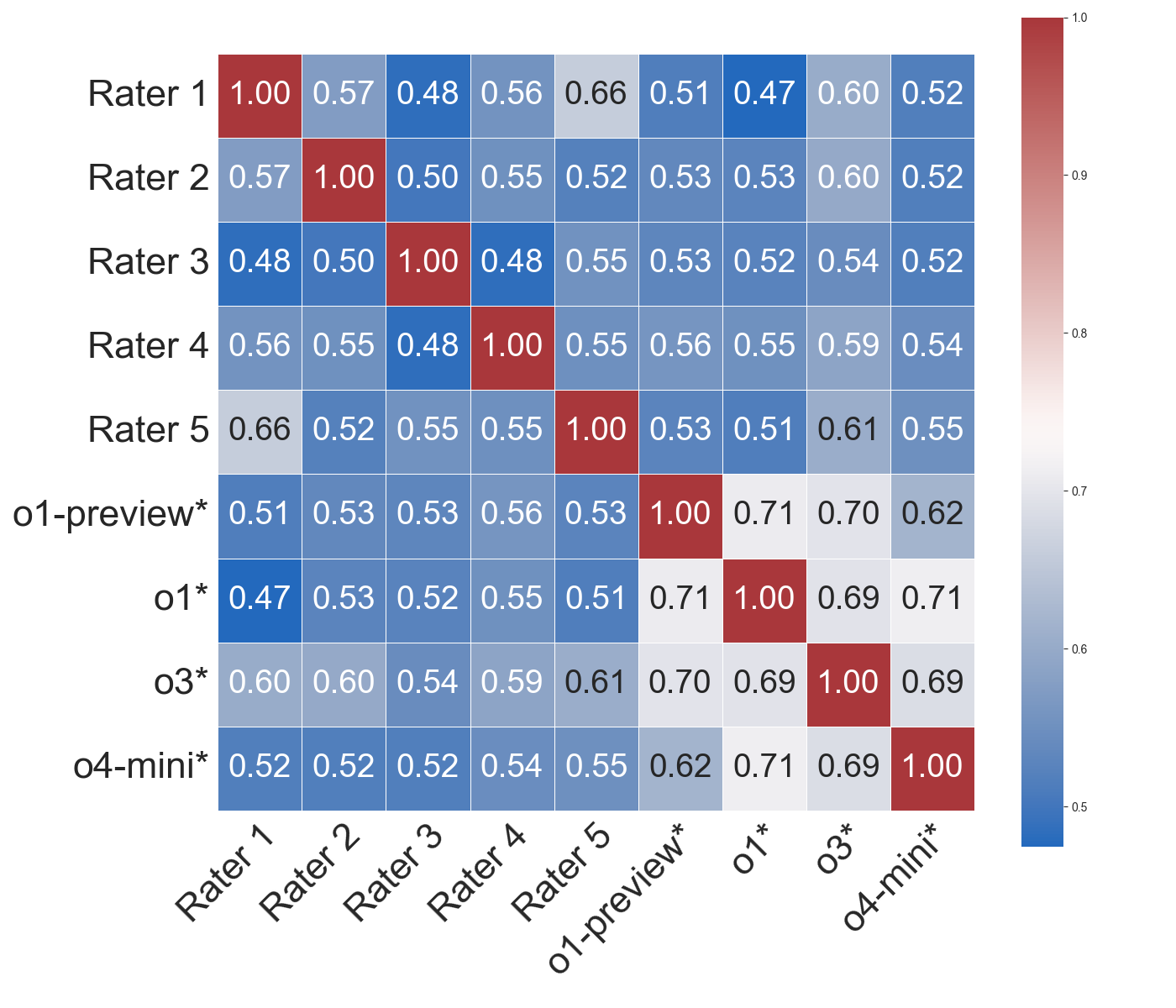}
\caption{Heatmap of pairwise Kendall's $\tau$ coefficients between assessors for the ranking task.
Values indicate the degree of agreement in ranking orders.
AI assessors are marked with asterisks.}
\label{fig:kendall_tau_heatmap}
\end{figure}

GPT-4o achieved the lowest average rank of 3.21 (median = 3.0), followed by GPT-3.5 Turbo with an average rank of 3.46 (median = 3.0).
Among open-source models, SauerkrautLM Mixtral 8×7b Q8 achieved an average rank of 4.26 (median = 4.0).
SauerkrautLM Mixtral 8×7b Q4 followed with mean = 4.95 (median = 5.0) and SauerkrautLM Llama 3 70b Q4 with mean = 5.19 (median = 4.0, std = 3.18).
The middle tier comprised Mixtral 8×7b Q4 (mean = 5.81, median = 6.0) and Mixtral 8×7b Q8 (mean = 6.63, median = 7.0).
Lower-tier models included SauerkrautLM Llama 3.1 8b Q8 (mean = 7.05, median = 9.0) through to Meta Llama 3.1 8b Q8 (mean = 9.27, median = 10.0).
Models with predominantly Good ratings (green dots) clustered in lower rank positions, whilst those with Poor ratings (red dots) concentrated in higher ranks.
Fair ratings (yellow dots) showed more dispersion across ranking positions.
Pairwise Kendall's $\tau$ coefficients were computed between all assessor pairs to measure ordinal association between rankings (Figure \ref{fig:kendall_tau_heatmap}).

Kendall's $\tau$ values ranged from 0.47 to 0.71 across all assessor pairs.
Human assessors (1–5) showed $\tau$ values from 0.48 to 0.66 (mean = 0.54).
The highest human–human agreement occurred between Rater 1 and Rater 5 ($\tau$ = 0.66).
The lowest occurred between Rater 3 and both Rater 1 and Rater 4 ($\tau$ = 0.48).
AI assessors demonstrated $\tau$ values between 0.62 and 0.71 (mean = 0.68).
The highest AI–AI agreement occurred between o1-preview and o1 ($\tau$ = 0.71) and between o1 and o4-mini ($\tau$ = 0.71).
Human–AI agreement ranged from 0.47 to 0.61 (mean = 0.54).
o3 showed the highest average agreement with human assessors ($\tau$ = 0.59), whilst o1 showed the lowest ($\tau$ = 0.52).
For eleven models, $\tau \ge 0.80$ indicates very high agreement, $0.60 \le \tau < 0.80$ represents good agreement, $0.40 \le \tau < 0.60$ shows moderate agreement and $\tau < 0.40$ indicates low agreement.
All observed values fell within the moderate to good agreement range.
No negative values were observed and the standard deviation of $\tau$ was approximately 0.07.

\subsection{Correlation Between Rating and Ranking Methods}
\label{sec: correlation_analysis}

\begin{figure}[htb]
\centering
\includegraphics[width=\linewidth]{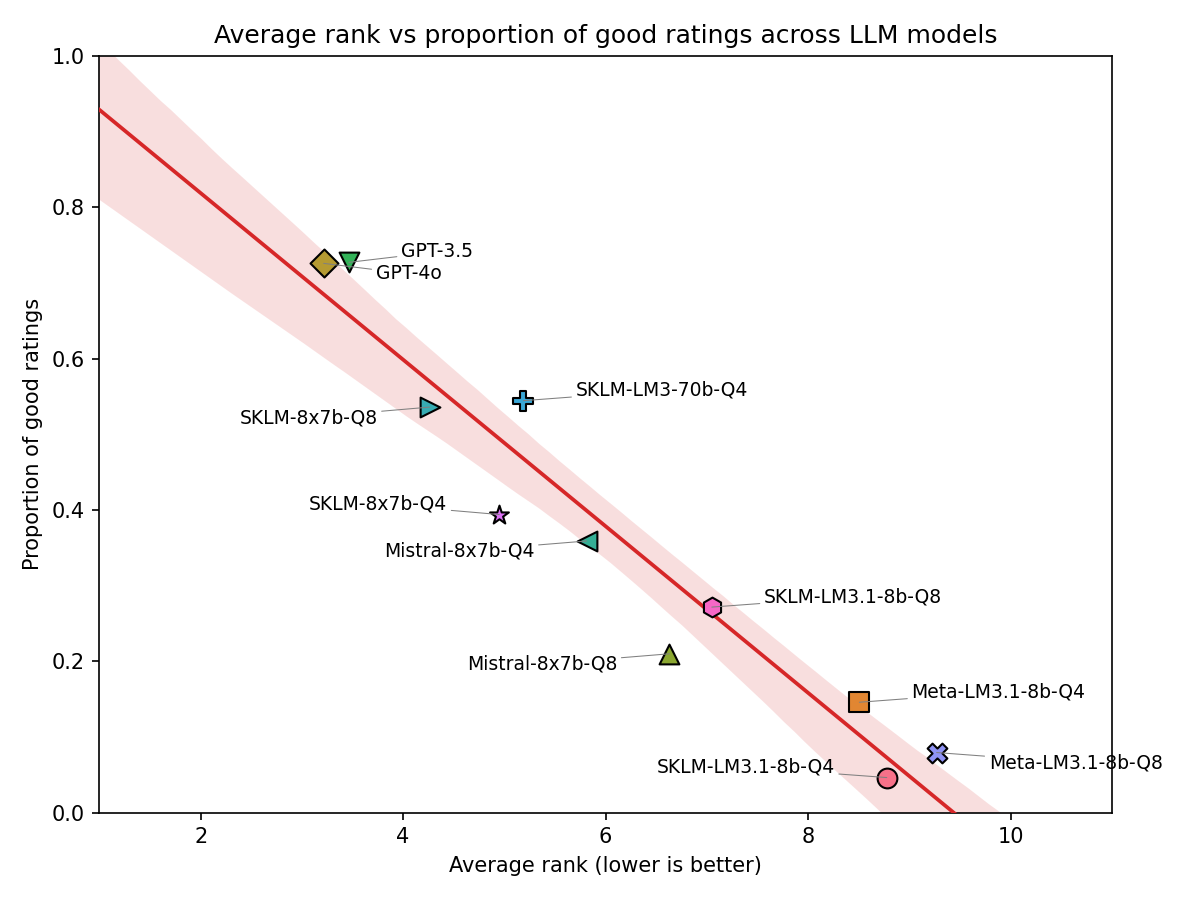}
\caption{Average rank versus proportion of Good ratings for all evaluated models. The red line represents the least-squares regression fit with 95\% confidence interval (shaded area). Models above the line achieve higher Good ratings than their rank, whilst those below achieve fewer Good ratings than their rank.}
\label{fig:rating_ranking_correlation}
\end{figure}

Linear regression analysis yielded the relationship $y = -0.1078x + 0.9957$, where $y$ represents the proportion of Good ratings and $x$ represents the average rank (see Figure~\ref{fig:rating_ranking_correlation}).
The coefficient of determination was $R^2 = 0.9499$.
Pearson correlation was $r = -0.9746$ and Spearman correlation was $\rho = -0.9545$.
Residual analysis showed models positioned at varying distances from the regression line.

The plot shows a clear negative association, models with lower average ranks tend to receive higher proportions of Good ratings. 
Points above the regression line outperform what their rank would predict. 
This group includes OpenAI GPT-4o and OpenAI GPT-3.5 (average ranks 3.21 and 3.46; both 73\% Good), as well as SauerkrautLM Llama 3 70b Q4 (5.19; 54\%). 
SauerkrautLM Mixtral 8×7b Q8 sits slightly below the line at 54\% Good with an average rank of 4.26, indicating marginal underperformance relative to its rank. SauerkrautLM Mixtral 8×7b Q4 (4.95; 39\%) and the Mistral 8×7b variants (5.81; 36\% and 6.63; 21\%) also fall below expectations. 
Meta Llama 3.1 8b Q8 (9.27; 8\%), Meta Llama 3.1 8b Q4 (8.51; 15\%), SauerkrautLM Llama 3.1 8b Q8 (7.05; 27\%) and SauerkrautLM Llama 3.1 8b Q4 (8.78; 5\%) all lie close to the regression fit. 
Overall, the upper-left quadrant is populated by strong performers, whilst the lower-right quadrant contains models with both high mean ranks and low Good proportions. 
Despite identical 73\% Good shares, OpenAI GPT-4o attains the better average rank compared with OpenAI GPT-3.5 Turbo.

\section{Discussion}
\label{sec: overall_performance}

The comprehensive hierarchical assessment provides converging evidence that addresses the fundamental research questions about LLM capabilities in generating subject lines for psychosocial counselling.
By synthesising insights from categorical ratings, ranking assessments and their correlation, this section examines model performance across three key dimensions.

\subsection{Performance Hierarchy: LLM Capability for Subject Line Generation}
This expanded study confirms the initial findings \cite{steigerwald_comparing_2025} that LLMs can effectively generate meaningful subject lines for psychosocial counselling emails.
The highest-performing models, GPT-3.5 Turbo and GPT-4o, achieved 73\% Good ratings using only generic prompting without task-specific fine-tuning or coun\-selling-do\-main train\-ing.
These out-of-the-box results establish a strong baseline, demonstrating that general-purpose LLMs already possess sufficient capability for practical deployment in subject line generation.

The best open-source alternatives, SauerkrautLM Llama 3 70b Q4 and SauerkrautLM Mixtral 8×7b Q8, achieved 54\% Good ratings\---a 19-percentage-point gap from proprietary leaders.
Whilst this gap appears substantial, several factors suggest it could be narrowed or eliminated through targeted optimisation.
The German-tuned SauerkrautLM variants already demonstrate significant improvements over base models, with SauerkrautLM Mixtral 8×7b Q8 outperforming its base counterpart by 33 percentage points (54\% vs 21\% Good ratings).

The relationship between categorical ratings and ranking assessments proved strongly linear, as demonstrated by the regression equation $y = -0.1078x + 0.9957$ ($R^2 = 0.95$), where $y$ represents Good rating proportions and $x$ represents average rank.
This near-perfect linear relationship (Pearson $r = -0.9746$, Spearman $\rho = -0.9545$) confirms that both methods measure the same underlying quality construct.
The strong correlation was expected given that rankings occurred within established categories, yet it raises the question of whether the additional ranking effort yields meaningful discrimination beyond categorical assessment.

The hierarchical design\---categorising first, then ranking within tiers\---transformed an overwhelming eleven-way comparison into manageable tasks for human assessors.
This approach proved essential. 
Whilst direct ranking of all models simultaneously would have been cognitively prohibitive, the pre-categorisation made systematic evaluation feasible.
The key question was whether this additional ranking step would reveal substantial performance differences masked by categorical ratings.

Whilst GPT-3.5 Turbo and GPT-4o both achieved 73\% Good ratings, ranking distinguished GPT-4o's consistent superiority (average rank 3.21 versus 3.46).
Similarly, median ranks of 3.0 for proprietary systems versus 4.0 for the best open-source alternatives provide granularity beyond categorical percentages.
However, these refinements represent incremental rather than transformative insights\---the ceiling effect exists but remains limited to the highest performance tier, suggesting that categorical ratings alone suffice for most practical deployment decisions.

Performance variation across the eleven models ranged from 73\% to 5\% Good ratings, indicating that successful subject line generation requires specific model capabilities.
The stark performance differences between SauerkrautLM variants and their base models suggest that domain-specific fine-tuning plays a crucial role.
Further prompt engineering, task-specific training, or increased model scale could potentially enable open-source models to match or exceed proprietary performance whilst maintaining the privacy advantages of local deployment.

The consistency of these findings with the original study \cite{steigerwald_comparing_2025}, now validated with an expanded assessor panel including multiple AI systems, strengthens confidence in LLM capabilities for this specific counselling support task.

\subsection{German Language Adaptation: SauerkrautLM vs Base Models}

The expanded study confirms and strengthens the original findings \cite{steigerwald_comparing_2025} regarding the substantial benefits of German language fine-tuning for counselling applications.
German-tuned SauerkrautLM variants outperformed their base counterparts across both categorical and ranking assessments.

The most dramatic improvement appeared in the Mixtral 8×7b architecture, where SauerkrautLM Mixtral 8×7b Q8 achieved 54\% Good ratings compared to the base model's 21\%\---a 33-percentage-point enhancement.
This represents a 2.6-fold improvement in Good rating proportions through German adaptation alone.
The Q4 variant showed a more modest but still meaningful improvement, with SauerkrautLM Mixtral 8×7b Q4 reaching 39\% Good ratings versus 36\% for the base model.
SauerkrautLM Llama 3 70b Q4 demonstrated that German fine-tuning can elevate open-source models to competitive performance levels, achieving 54\% Good ratings\---matching SauerkrautLM Mixtral 8×7b Q8 and approaching the proprietary models' performance.
This finding validates the potential of language-specific adaptation to bridge the quality gap between open-source and proprietary systems.

Ranking analysis reinforced these patterns with even clearer distinctions.
SauerkrautLM Mixtral 8×7b Q8 achieved a median rank of 4.0, substantially outperforming the base Mixtral 8×7b Q8's median rank of 7.0\---a three-rank improvement.


The sole exception occurred with the smallest model architecture: SauerkrautLM Llama 3.1 8b Q4 achieved only 5\% Good ratings compared to 15\% for the base Meta Llama 3.1 8b Q4.
Both models performed poorly overall, indicating that German fine-tuning cannot compensate for insufficient baseline model capacity.

The consistent superiority of German-tuned models across multiple architectures, quantisation levels and evaluation methods establishes language-specific fine-tuning as a critical factor for counselling applications.
The improvements likely stem from better comprehension of German linguistic nuances, counselling terminology and culturally specific expressions of psychological distress.

\subsection{Model Architecture: Parameter Count and Quantisation Effects}

Model size demonstrated a strong correlation with subject line generation quality.
The evaluation established a clear hierarchy, larger models generally outperformed smaller ones across both assessment methods.

The top-performing models, GPT-4o and GPT-3.5 Turbo, both achieved 73\% Good ratings with median ranks of 3.0.
Whilst GPT-4o's exact parameter count remains undisclosed, estimates suggest hundreds of billions of parameters.
GPT-3.5 Turbo, with an estimated 20b parameters, matched this performance despite being substantially smaller\---indicating that architectural efficiency and training quality can compensate for raw parameter count.
The 70b-parameter SauerkrautLM Llama 3 70b Q4 achieved 54\% Good ratings, demonstrating that open-source models at this scale can deliver practical utility.
The Mixtral architecture, with 47b total parameters but only 13b active per expert through mixture-of-experts design, showed the widest performance variation (21\% to 54\% Good ratings), suggesting that model architecture interacts with language adaptation.
The 8b-parameter models consistently underperformed, achieving only 5\% to 27\% Good ratings and clustering in median ranks 7–11.
This finding establishes a practical threshold, models below approximately 13b active parameters struggle to generate quality subject lines for counselling contexts, regardless of fine-tuning efforts.

Quantisation effects proved inconsistent and model-specific rather than following predictable patterns.
Within the SauerkrautLM Mixtral variants, the Q8 version achieved 54\% Good ratings whilst Q4 reached only 39\%\---suggesting that 8-bit precision preserves crucial capabilities for this model.
Conversely, the base Mixtral showed the opposite pattern: Q4 achieved 36\% Good ratings versus Q8's 21\%.
For the smaller Llama 3.1 8b architecture, quantisation showed minimal impact on already poor performance.
Both Q4 and Q8 variants of the base model achieved similarly low Good ratings (15\% and 8\% respectively), whilst the SauerkrautLM versions performed poorly regardless of quantisation (5\% for Q4, 27\% for Q8).

These findings indicate that quantisation interacts with model architecture, size and language adaptation.
The absence of systematic quality degradation from 8-bit to 4-bit quantisation for certain models presents opportunities for resource-constrained deployments.
Organisations could potentially halve memory requirements using 4-bit models without compromising quality, though model-specific evaluation remains essential given the unpredictable interaction effects.

The performance patterns suggest a minimum viable configuration for counselling applications: at least 13b active parameters with task-appropriate fine-tuning.
Below this threshold, neither architectural innovations nor language adaptation can compensate for insufficient model capacity.

\subsection{Evaluation Methodology: Human vs AI Assessors}

The expanded assessor panel\---five human counselling professionals and four AI systems\---revealed systematic differences in evaluation consistency.

AI assessors demonstrated higher inter-rater agreement across both methods.
For categorical ratings, AI systems achieved mean Spearman correlation of 0.78, whilst human assessors showed 0.70.
In ranking tasks, AI assessors achieved mean Kendall's $\tau$ of 0.69, compared to 0.54 for human assessors.
Human-AI agreement fell between these extremes at 0.69 (categorical) and 0.54 (ranking).

The variability in human assessment represents crucial strength rather than limitation.
Human assessors remain the gold standard for evaluating counselling communications because their diverse professional experiences capture the full spectrum of quality considerations.
Each counselling professional brings unique clinical perspectives, ensuring subject lines are evaluated for counselling appropriateness across varied contexts, not just technical correctness.

The hierarchical assessment design proved essential for maintaining human evaluation quality.
By reducing cognitive load through structured steps\---first categorising then ranking\---the framework enabled reliable expert judgments without overwhelming complexity.
This methodological simplicity becomes critical, the simpler the assessment task, the more reliable human expert judgments become.

AI assessors demonstrated value as supplementary evaluators, providing consistent baseline measurement particularly suited for initial screening.
However, moderate human-AI agreement indicates that AI systems apply different quality criteria than human experts.
For counselling applications requiring empathy, cultural sensitivity and counselling appropriateness, human evaluation must remain primary.

The current framework positions AI assessment appropriately/---as enrichment to human expertise rather than replacement.
This complementary approach\---human expertise as foundation with AI enrichment\---validates that whilst AI offers reliable supplementary assessment, counselling professionals' nuanced judgment remains irreplaceable for mental health communication.
Future frameworks should maintain this hierarchy through simplified, structured methodologies for human experts, with AI systems providing additional consistency checks.

\subsection{Hierarchical Assessment: Categorical Ratings vs Rankings}

The hierarchical assessment framework combined categorical ratings with ranking analysis to examine the extent of ceiling effects in categorical assessment within this counselling context.
The strong correlation between methods (Pearson $r = -0.9746$, Spearman $\rho = -0.9545$) was expected given that rankings occurred within established categories.
Analysis revealed ceiling effects primarily at the top performance tier where GPT-3.5 Turbo and GPT-4o both achieved 73\% Good ratings.
Pure categorical assessment could not differentiate between these models.
Ranking analysis resolved this limitation, suggesting GPT-4o's slightly superiority (average rank 3.21 vs 3.46).
Similarly, among models achieving 54\% Good ratings, ranking distinguished SauerkrautLM Mixtral 8×7b Q8 (rank 4.26) from SauerkrautLM Llama 3 70b Q4 (rank 5.19).

The ceiling effect in this dataset manifested in two performance clusters (73\% and 54\%) containing multiple models requiring ranking discrimination.
The remaining models distributed across the full performance spectrum from 39\% down to 5\% Good ratings, indicating that categorical assessment retained discriminative power for most comparisons in this counselling application.
Beyond resolving ceiling effects, the ranking process provided methodological benefits.
Requiring assessors to make relative comparisons encouraged deeper engagement with each subject line's qualities.
Rather than evaluating outputs in isolation, assessors considered subtle differences in clarity, specificity and counselling appropriateness when ordering models within categories.
This comparative process likely enhanced assessment quality, particularly for human assessors drawing on clinical experience to distinguish nuanced differences.

In sum, the hierarchical design suggests that, for this counselling task, categorical ratings are generally sufficient and lighten the burden on human assessors. A ranking analysis need only be introduced as a second step when LLM performances are tightly clustered and ceiling effects emerge, serving as a targeted tie-breaker rather than a default requirement.

\section{Ethical Considerations}
The deployment of LLMs for subject-line generation in psychosocial counselling demands careful ethical examination.

Privacy remains paramount\---GPT-4o and GPT-3.5 Turbo (73\% Good) require transmitting sensitive data to external servers, whilst SauerkrautLM variants (54\% Good) enable local deployment.
This 19-percentage-point gap creates tension between privacy protection and service effectiveness.
Fairness considerations emerge from the evaluation methodology.
The filtering process may have excluded marginalised experiences or non-normative expressions of distress.
High-retention models might excel at mainstream scenarios whilst failing culturally specific communications.
Accountability frameworks must delineate responsibility when automatically generated subject lines misrepresent client needs, potentially delaying critical interventions.
Transparency must ensure that counsellors are fully aware when AI is involved in generating subject lines and that they are adequately trained to interpret and critically evaluate these outputs.
This includes understanding both the capabilities and limitations of the technology, so that human expertise remains central to decision-making.
Beneficence supports the technology\---transforming vague subject lines could improve response times and care quality.
Non-maleficence demands acknowledging that even top models produce 27\% non-Good outputs, potentially including harmful misrepresentations.
Justice principles highlight systemic inequities.
Expensive proprietary models create two-tier systems, whilst viable open-source alternatives could democratise AI-enhanced counselling.
Four-bit quantisation enabling deployment without quality loss promotes equity by lowering computational barriers.
Environmental justice emerges\---SauerkrautLM Mixtral 8×7b Q8 uses substantially fewer resources than GPT-4o whilst maintaining practical utility.

These considerations underscore that ethical guidelines are not optional additions but fundamental requirements.
Technology deployment in mental health cannot proceed through trial and error\---the stakes are too high and the population too vulnerable.
Before implementation, systematic consideration must examine how AI tools reshape counselling dynamics, alter counselling relationships and potentially transform help-seeking behaviours.
Long-term consequences require careful anticipation: will AI-generated subject lines inadvertently train counsellors to expect pre-digested information?
Might clients adapt their writing to optimise for AI processing rather than authentic expression?

Responsible development demands proactive safeguards from inception, not reactive patches after harm emerges.
Systems must incorporate adjustable parameters\---confidence thresholds, human-override mechanisms, bias detection triggers\---enabling continuous refinement as understanding evolves.
The path forward requires treating ethical considerations as design constraints rather than deployment obstacles, ensuring that efficiency gains never compromise the fundamental trust and human connection at counselling's core.

\section{Conclusion}

This expanded study confirms that Large Language Models can generate meaningful subject lines for psychosocial counselling contexts.
GPT-4o and GPT-3.5 Turbo achieved 73\% Good ratings, whilst the best open-source alternatives, SauerkrautLM Llama 3 70b Q4 and SauerkrautLM Mixtral 8×7b Q8, reached 54\%.
German-tuned models outperformed base versions by up to 33 percentage points, demonstrating the critical importance of language-specific adaptation.

The 19-percentage-point performance gap can likely be closed through targeted optimisation.
Future refinements\---including fine-tuning with counselling-specific subject lines and reinforcement learning approaches such as Direct Preference Optimization (DPO)\---could enable open-source models to match proprietary performance.
Combined with robust ethical guidelines, these improvements should make AI-powered subject line generation viable for routine e-mental health deployment.
However, user acceptance and actual clinical benefit require systematic investigation through real-world pilot studies.

The hierarchical assessment approach proved effective, with pre-categorisation enabling otherwise infeasible ranking tasks.
However, the strong correlation between methods demonstrates that categorical ratings alone provide sufficient discrimination for practical deployment decisions.
Human evaluation remains the gold standard whilst AI assessors provide valuable consistency checks, validating the assessment methodology.
Privacy-preserving deployment through open-source models proves viable, with SauerkrautLM variants demonstrating effective generation whilst maintaining data sovereignty\---crucial for European data-protection compliance.

Integrating LLMs into psychosocial counselling represents both tremendous opportunity and significant responsibility.
Whilst subject line generation represents merely one small support function, these technologies could substantially reduce counsellor workload through various applications\---from case summarisation and documentation assistance to pattern recognition across client communications.
By automating time-consuming administrative tasks that LLMs can effectively handle, counselling professionals could dedicate more time to direct counselling work, potentially addressing rising service demand without compromising care quality.
Realising this potential demands unwavering commitment to rigorous evaluation, ethical deployment and continuous improvement.

The path forward is clear.
When ethical principles guide development, regulatory frameworks ensure accountability and quality standards remain uncompromising, LLMs can genuinely enhance e-mental-health services.
By consciously steering this technological evolution with human welfare at its centre, the field can harness AI's capabilities whilst preserving the trust and confidentiality essential to effective counselling.
The journey toward AI-augmented mental-health support has begun.
With careful stewardship, sustained research and principled implementation, these technologies can become powerful tools for extending mental-health assistance to those who need it most.

\bibliographystyle{splncs04}
\bibliography{clean_references_ICT4AWE_PostPub}

@article{ierardi_effectiveness_2022,
	title = {Effectiveness of an online versus face-to-face psychodynamic counselling intervention for university students before and during the {COVID}-19 period},
	volume = {10},
	issn = {2050-7283},	number = {1},	journal = {BMC Psychology},
	author = {Ierardi, Elena and Bottini, Marta and Riva Crugnola, Cristina},
	month = feb,
	year = {2022},	pages = {35},
}

@article{aspvall_effect_2021,
	title = {Effect of an {Internet}-{Delivered} {Stepped}-{Care} {Program} vs {In}-{Person} {Cognitive} {Behavioral} {Therapy} on {Obsessive}-{Compulsive} {Disorder} {Symptoms} in {Children} and {Adolescents}: {A} {Randomized} {Clinical} {Trial}},
	volume = {325},
	issn = {0098-7484},	number = {18},	journal = {JAMA},
	author = {Aspvall, Kristina and Andersson, Erik and Melin, Karin and Norlin, Lisa and Eriksson, Viktor and Vigerland, Sarah and Jolstedt, Maral and Silverberg-Mörse, Maria and Wallin, Lena and Sampaio, Filipa and Feldman, Inna and Bottai, Matteo and Lenhard, Fabian and Mataix-Cols, David and Serlachius, Eva},
	month = may,
	year = {2021},
	pages = {1863--1873},
}

@article{lattie_overview_2022,
	title = {An overview of and recommendations for more accessible digital mental health services},
	volume = {1},	issn = {2731-0574},	number = {2},	journal = {Nature Reviews Psychology},
	author = {Lattie, Emily G. and Stiles-Shields, Colleen and Graham, Andrea K.},
	month = feb,
	year = {2022},	pages = {87--100},
}

@article{guo_large_2024,
	title = {Large {Language} {Models} for {Mental} {Health} {Applications}: {Systematic} {Review}},
	volume = {11},	number = {1},	journal = {JMIR Mental Health},
	author = {Guo, Zhijun and Lai, Alvina and Thygesen, Johan H. and Farrington, Joseph and Keen, Thomas and Li, Kezhi},
	month = oct,
	year = {2024},	pages = {e57400},
}

@inproceedings{balloccu_ask_2024,	title = {Ask the experts: sourcing a high-quality nutrition counseling dataset through {Human}-{AI} collaboration},	booktitle = {Findings of the {Association} for {Computational} {Linguistics}: {EMNLP} 2024},
	publisher = {Association for Computational Linguistics},
	author = {Balloccu, Simone and Reiter, Ehud and Li, Karen Jia-Hui and Sargsyan, Rafael and Kumar, Vivek and Reforgiato, Diego and Riboni, Daniele and Dusek, Ondrej},
	editor = {Al-Onaizan, Yaser and Bansal, Mohit and Chen, Yun-Nung},
	month = nov,
	year = {2024},
	pages = {11519--11545},
}

@inproceedings{liusie_llm_2024,	title = {{LLM} {Comparative} {Assessment}: {Zero}-shot {NLG} {Evaluation} through {Pairwise} {Comparisons} using {Large} {Language} {Models}},	booktitle = {Proceedings of the 18th {Conference} of the {European} {Chapter} of the {Association} for {Computational} {Linguistics} ({Volume} 1: {Long} {Papers})},
	publisher = {Association for Computational Linguistics},
	author = {Liusie, Adian and Manakul, Potsawee and Gales, Mark},
	editor = {Graham, Yvette and Purver, Matthew},
	month = mar,
	year = {2024},
	pages = {139--151},
}

@inproceedings{belz_comparing_2010,
	title = {Comparing {Rating} {Scales} and {Preference} {Judgements} in {Language} {Evaluation}},	booktitle = {Proceedings of the 6th {International} {Natural} {Language} {Generation} {Conference}},
	publisher = {Association for Computational Linguistics},
	author = {Belz, Anja and Kow, Eric},
	editor = {Kelleher, John and Namee, Brian Mac and Sluis, Ielka van der},
	month = jul,
	year = {2010},
}

@misc{liu_g-eval_2023,
	title = {G-{Eval}: {NLG} {Evaluation} using {GPT}-4 with {Better} {Human} {Alignment}},	publisher = {arXiv},
	author = {Liu, Yang and Iter, Dan and Xu, Yichong and Wang, Shuohang and Xu, Ruochen and Zhu, Chenguang},
	month = may,
	year = {2023},}

@misc{gao_human-like_2023,
	title = {Human-like {Summarization} {Evaluation} with {ChatGPT}},	publisher = {arXiv},
	author = {Gao, Mingqi and Ruan, Jie and Sun, Renliang and Yin, Xunjian and Yang, Shiping and Wan, Xiaojun},
	month = apr,
	year = {2023},}

@inproceedings{thomson_mostly_2024,	title = {({Mostly}) {Automatic} {Experiment} {Execution} for {Human} {Evaluations} of {NLP} {Systems}},	booktitle = {Proceedings of the 17th {International} {Natural} {Language} {Generation} {Conference}},
	publisher = {Association for Computational Linguistics},
	author = {Thomson, Craig and Belz, Anya},
	editor = {Mahamood, Saad and Minh, Nguyen Le and Ippolito, Daphne},
	month = sep,
	year = {2024},
	pages = {272--279},
}

@misc{hsu_helping_2023,
	title = {Helping the {Helper}: {Supporting} {Peer} {Counselors} via {AI}-{Empowered} {Practice} and {Feedback}},	publisher = {arXiv},
	author = {Hsu, Shang-Ling and Shah, Raj Sanjay and Senthil, Prathik and Ashktorab, Zahra and Dugan, Casey and Geyer, Werner and Yang, Diyi},
	month = may,
	year = {2023},}

@inproceedings{rudolph_ai-based_2024,	title = {An {AI}-{Based} {Virtual} {Client} for {Educational} {Role}-{Playing} in the {Training} of {Online} {Counselors}:},	booktitle = {Proceedings of the 16th {International} {Conference} on {Computer} {Supported} {Education}},
	publisher = {SCITEPRESS - Science and Technology Publications},
	author = {Rudolph, Eric and Engert, Natalie and Albrecht, Jens},
	year = {2024},	pages = {108--117},
}

@misc{fu_enhancing_2023,
	title = {Enhancing {Psychological} {Counseling} with {Large} {Language} {Model}: {A} {Multifaceted} {Decision}-{Support} {System} for {Non}-{Professionals}},	publisher = {arXiv},
	author = {Fu, Guanghui and Zhao, Qing and Li, Jianqiang and Luo, Dan and Song, Changwei and Zhai, Wei and Liu, Shuo and Wang, Fan and Wang, Yan and Cheng, Lijuan and Zhang, Juan and Yang, Bing Xiang},
	month = aug,
	year = {2023},}

@inproceedings{steigerwald_comparing_2025,	title = {Comparing {Large} {Language} {Models} for {Automated} {Subject} {Line} {Generation} in e-{Mental} {Health}: {A} {Performance} {Study}:},	booktitle = {Proceedings of the 11th {International} {Conference} on {Information} and {Communication} {Technologies} for {Ageing} {Well} and e-{Health}},
	publisher = {SCITEPRESS - Science and Technology Publications},
	author = {Steigerwald, Philipp and Albrecht, Jens},
	year = {2025},
	pages = {70--77},
}

@article{koutsouleris_promise_2022,
	title = {From promise to practice: towards the realisation of {AI}-informed mental health care},
	volume = {4},
	issn = {2589-7500},	number = {11},	journal = {The Lancet Digital Health},
	author = {Koutsouleris, Nikolaos and Hauser, Tobias U. and Skvortsova, Vasilisa and Choudhury, Munmun De},
	month = nov,
	year = {2022},
	pmid = {36229346},	pages = {e829--e840},
}

@misc{chiu_computational_2024,
	title = {A {Computational} {Framework} for {Behavioral} {Assessment} of {LLM} {Therapists}},	publisher = {arXiv},
	author = {Chiu, Yu Ying and Sharma, Ashish and Lin, Inna Wanyin and Althoff, Tim},
	month = jan,
	year = {2024},}

@misc{liu_chatcounselor_2023,
	title = {{ChatCounselor}: {A} {Large} {Language} {Models} for {Mental} {Health} {Support}},	publisher = {arXiv},
	author = {Liu, June M. and Li, Donghao and Cao, He and Ren, Tianhe and Liao, Zeyi and Wu, Jiamin},
	month = sep,
	year = {2023},}

@inproceedings{steigerwald_enhancing_2024,	title = {Enhancing {Psychosocial} {Counselling} with {AI}: {A} {Multifaceted} {Support} {System} for {Professionals}},	booktitle = {Frontiers of {Artificial} {Intelligence}, {Ethics}, and {Multidisciplinary} {Applications}},
	publisher = {Springer Nature},
	author = {Steigerwald, Philipp and Albrecht, Jens},
	month = oct,
	year = {2024},}

@article{vowels_ai_2024,
	title = {{AI} in relationship counselling: {Evaluating} {ChatGPT}'s therapeutic capabilities in providing relationship advice},
	volume = {2},
	issn = {2949-8821},	number = {2},	journal = {Computers in Human Behavior: Artificial Humans},
	author = {Vowels, Laura M. and Francois-Walcott, Rachel R. R. and Darwiche, Joëlle},
	month = aug,
	year = {2024},	pages = {100078},
}

@article{lai_supporting_2024,
	title = {Supporting the {Demand} on {Mental} {Health} {Services} with {AI}-{Based} {Conversational} {Large} {Language} {Models} ({LLMs})},
	volume = {4},	issn = {2673-7426},	number = {1},	journal = {BioMedInformatics},
	author = {Lai, Tin and Shi, Yukun and Du, Zicong and Wu, Jiajie and Fu, Ken and Dou, Yichao and Wang, Ziqi},
	month = mar,
	year = {2024},	pages = {8--33},
}

@misc{dubey_llama_2024,
	title = {The {Llama} 3 {Herd} of {Models}},	publisher = {arXiv},
	author = {Dubey, Abhimanyu and Jauhri, Abhinav and Pandey, Abhinav and Kadian, Abhishek and Al-Dahle, Ahmad and Letman, Aiesha and Mathur, Akhil and Schelten, Alan and Yang, Amy and Fan, Angela and Goyal, Anirudh and Hartshorn, Anthony and Yang, Aobo and Mitra, Archi and Sravankumar, Archie and Korenev, Artem and Hinsvark, Arthur and Rao, Arun and Zhang, Aston and Rodriguez, Aurelien and Gregerson, Austen and Spataru, Ava and Roziere, Baptiste and Biron, Bethany and Tang, Binh and Chern, Bobbie and Caucheteux, Charlotte and Nayak, Chaya and Bi, Chloe and Marra, Chris and McConnell, Chris and Keller, Christian and Touret, Christophe and Wu, Chunyang and Wong, Corinne and Ferrer, Cristian Canton and Nikolaidis, Cyrus and Allonsius, Damien and Song, Daniel and Pintz, Danielle and Livshits, Danny and Esiobu, David and Choudhary, Dhruv and Mahajan, Dhruv and Garcia-Olano, Diego and Perino, Diego and Hupkes, Dieuwke and Lakomkin, Egor and AlBadawy, Ehab and Lobanova, Elina and Dinan, Emily and Smith, Eric Michael and Radenovic, Filip and Zhang, Frank and Synnaeve, Gabriel and Lee, Gabrielle and Anderson, Georgia Lewis and Nail, Graeme and Mialon, Gregoire and Pang, Guan and Cucurell, Guillem and Nguyen, Hailey and Korevaar, Hannah and Xu, Hu and Touvron, Hugo and Zarov, Iliyan and Ibarra, Imanol Arrieta and Kloumann, Isabel and Misra, Ishan and Evtimov, Ivan and Copet, Jade and Lee, Jaewon and Geffert, Jan and Vranes, Jana and Park, Jason and Mahadeokar, Jay and Shah, Jeet and Linde, Jelmer van der and Billock, Jennifer and Hong, Jenny and Lee, Jenya and Fu, Jeremy and Chi, Jianfeng and Huang, Jianyu and Liu, Jiawen and Wang, Jie and Yu, Jiecao and Bitton, Joanna and Spisak, Joe and Park, Jongsoo and Rocca, Joseph and Johnstun, Joshua and Saxe, Joshua and Jia, Junteng and Alwala, Kalyan Vasuden and Upasani, Kartikeya and Plawiak, Kate and Li, Ke and Heafield, Kenneth and Stone, Kevin and El-Arini, Khalid and Iyer, Krithika and Malik, Kshitiz and Chiu, Kuenley and Bhalla, Kunal and Rantala-Yeary, Lauren and Maaten, Laurens van der and Chen, Lawrence and Tan, Liang and Jenkins, Liz and Martin, Louis and Madaan, Lovish and Malo, Lubo and Blecher, Lukas and Landzaat, Lukas and Oliveira, Luke de and Muzzi, Madeline and Pasupuleti, Mahesh and Singh, Mannat and Paluri, Manohar and Kardas, Marcin and Oldham, Mathew and Rita, Mathieu and Pavlova, Maya and Kambadur, Melanie and Lewis, Mike and Si, Min and Singh, Mitesh Kumar and Hassan, Mona and Goyal, Naman and Torabi, Narjes and Bashlykov, Nikolay and Bogoychev, Nikolay and Chatterji, Niladri and Duchenne, Olivier and Çelebi, Onur and Alrassy, Patrick and Zhang, Pengchuan and Li, Pengwei and Vasic, Petar and Weng, Peter and Bhargava, Prajjwal and Dubal, Pratik and Krishnan, Praveen and Koura, Punit Singh and Xu, Puxin and He, Qing and Dong, Qingxiao and Srinivasan, Ragavan and Ganapathy, Raj and Calderer, Ramon and Cabral, Ricardo Silveira and Stojnic, Robert and Raileanu, Roberta and Girdhar, Rohit and Patel, Rohit and Sauvestre, Romain and Polidoro, Ronnie and Sumbaly, Roshan and Taylor, Ross and Silva, Ruan and Hou, Rui and Wang, Rui and Hosseini, Saghar and Chennabasappa, Sahana and Singh, Sanjay and Bell, Sean and Kim, Seohyun Sonia and Edunov, Sergey and Nie, Shaoliang and Narang, Sharan and Raparthy, Sharath and Shen, Sheng and Wan, Shengye and Bhosale, Shruti and Zhang, Shun and Vandenhende, Simon and Batra, Soumya and Whitman, Spencer and Sootla, Sten and Collot, Stephane and Gururangan, Suchin and Borodinsky, Sydney and Herman, Tamar and Fowler, Tara and Sheasha, Tarek and Georgiou, Thomas and Scialom, Thomas and Speckbacher, Tobias and Mihaylov, Todor and Xiao, Tong and Karn, Ujjwal and Goswami, Vedanuj and Gupta, Vibhor and Ramanathan, Vignesh and Kerkez, Viktor and Gonguet, Vincent and Do, Virginie and Vogeti, Vish and Petrovic, Vladan and Chu, Weiwei and Xiong, Wenhan and Fu, Wenyin and Meers, Whitney and Martinet, Xavier and Wang, Xiaodong and Tan, Xiaoqing Ellen and Xie, Xinfeng and Jia, Xuchao and Wang, Xuewei and Goldschlag, Yaelle and Gaur, Yashesh and Babaei, Yasmine and Wen, Yi and Song, Yiwen and Zhang, Yuchen and Li, Yue and Mao, Yuning and Coudert, Zacharie Delpierre and Yan, Zheng and Chen, Zhengxing and Papakipos, Zoe and Singh, Aaditya and Grattafiori, Aaron and Jain, Abha and Kelsey, Adam and Shajnfeld, Adam and Gangidi, Adithya and Victoria, Adolfo and Goldstand, Ahuva and Menon, Ajay and Sharma, Ajay and Boesenberg, Alex and Vaughan, Alex and Baevski, Alexei and Feinstein, Allie and Kallet, Amanda and Sangani, Amit and Yunus, Anam and Lupu, Andrei and Alvarado, Andres and Caples, Andrew and Gu, Andrew and Ho, Andrew and Poulton, Andrew and Ryan, Andrew and Ramchandani, Ankit and Franco, Annie and Saraf, Aparajita and Chowdhury, Arkabandhu and Gabriel, Ashley and Bharambe, Ashwin and Eisenman, Assaf and Yazdan, Azadeh and James, Beau and Maurer, Ben and Leonhardi, Benjamin and Huang, Bernie and Loyd, Beth and Paola, Beto De and Paranjape, Bhargavi and Liu, Bing and Wu, Bo and Ni, Boyu and Hancock, Braden and Wasti, Bram and Spence, Brandon and Stojkovic, Brani and Gamido, Brian and Montalvo, Britt and Parker, Carl and Burton, Carly and Mejia, Catalina and Wang, Changhan and Kim, Changkyu and Zhou, Chao and Hu, Chester and Chu, Ching-Hsiang and Cai, Chris and Tindal, Chris and Feichtenhofer, Christoph and Civin, Damon and Beaty, Dana and Kreymer, Daniel and Li, Daniel and Wyatt, Danny and Adkins, David and Xu, David and Testuggine, Davide and David, Delia and Parikh, Devi and Liskovich, Diana and Foss, Didem and Wang, Dingkang and Le, Duc and Holland, Dustin and Dowling, Edward and Jamil, Eissa and Montgomery, Elaine and Presani, Eleonora and Hahn, Emily and Wood, Emily and Brinkman, Erik and Arcaute, Esteban and Dunbar, Evan and Smothers, Evan and Sun, Fei and Kreuk, Felix and Tian, Feng and Ozgenel, Firat and Caggioni, Francesco and Guzmán, Francisco and Kanayet, Frank and Seide, Frank and Florez, Gabriela Medina and Schwarz, Gabriella and Badeer, Gada and Swee, Georgia and Halpern, Gil and Thattai, Govind and Herman, Grant and Sizov, Grigory and Guangyi and Zhang and Lakshminarayanan, Guna and Shojanazeri, Hamid and Zou, Han and Wang, Hannah and Zha, Hanwen and Habeeb, Haroun and Rudolph, Harrison and Suk, Helen and Aspegren, Henry and Goldman, Hunter and Damlaj, Ibrahim and Molybog, Igor and Tufanov, Igor and Veliche, Irina-Elena and Gat, Itai and Weissman, Jake and Geboski, James and Kohli, James and Asher, Japhet and Gaya, Jean-Baptiste and Marcus, Jeff and Tang, Jeff and Chan, Jennifer and Zhen, Jenny and Reizenstein, Jeremy and Teboul, Jeremy and Zhong, Jessica and Jin, Jian and Yang, Jingyi and Cummings, Joe and Carvill, Jon and Shepard, Jon and McPhie, Jonathan and Torres, Jonathan and Ginsburg, Josh and Wang, Junjie and Wu, Kai and U, Kam Hou and Saxena, Karan and Prasad, Karthik and Khandelwal, Kartikay and Zand, Katayoun and Matosich, Kathy and Veeraraghavan, Kaushik and Michelena, Kelly and Li, Keqian and Huang, Kun and Chawla, Kunal and Lakhotia, Kushal and Huang, Kyle and Chen, Lailin and Garg, Lakshya and A, Lavender and Silva, Leandro and Bell, Lee and Zhang, Lei and Guo, Liangpeng and Yu, Licheng and Moshkovich, Liron and Wehrstedt, Luca and Khabsa, Madian and Avalani, Manav and Bhatt, Manish and Tsimpoukelli, Maria and Mankus, Martynas and Hasson, Matan and Lennie, Matthew and Reso, Matthias and Groshev, Maxim and Naumov, Maxim and Lathi, Maya and Keneally, Meghan and Seltzer, Michael L. and Valko, Michal and Restrepo, Michelle and Patel, Mihir and Vyatskov, Mik and Samvelyan, Mikayel and Clark, Mike and Macey, Mike and Wang, Mike and Hermoso, Miquel Jubert and Metanat, Mo and Rastegari, Mohammad and Bansal, Munish and Santhanam, Nandhini and Parks, Natascha and White, Natasha and Bawa, Navyata and Singhal, Nayan and Egebo, Nick and Usunier, Nicolas and Laptev, Nikolay Pavlovich and Dong, Ning and Zhang, Ning and Cheng, Norman and Chernoguz, Oleg and Hart, Olivia and Salpekar, Omkar and Kalinli, Ozlem and Kent, Parkin and Parekh, Parth and Saab, Paul and Balaji, Pavan and Rittner, Pedro and Bontrager, Philip and Roux, Pierre and Dollar, Piotr and Zvyagina, Polina and Ratanchandani, Prashant and Yuvraj, Pritish and Liang, Qian and Alao, Rachad and Rodriguez, Rachel and Ayub, Rafi and Murthy, Raghotham and Nayani, Raghu and Mitra, Rahul and Li, Raymond and Hogan, Rebekkah and Battey, Robin and Wang, Rocky and Maheswari, Rohan and Howes, Russ and Rinott, Ruty and Bondu, Sai Jayesh and Datta, Samyak and Chugh, Sara and Hunt, Sara and Dhillon, Sargun and Sidorov, Sasha and Pan, Satadru and Verma, Saurabh and Yamamoto, Seiji and Ramaswamy, Sharadh and Lindsay, Shaun and Lindsay, Shaun and Feng, Sheng and Lin, Shenghao and Zha, Shengxin Cindy and Shankar, Shiva and Zhang, Shuqiang and Zhang, Shuqiang and Wang, Sinong and Agarwal, Sneha and Sajuyigbe, Soji and Chintala, Soumith and Max, Stephanie and Chen, Stephen and Kehoe, Steve and Satterfield, Steve and Govindaprasad, Sudarshan and Gupta, Sumit and Cho, Sungmin and Virk, Sunny and Subramanian, Suraj and Choudhury, Sy and Goldman, Sydney and Remez, Tal and Glaser, Tamar and Best, Tamara and Kohler, Thilo and Robinson, Thomas and Li, Tianhe and Zhang, Tianjun and Matthews, Tim and Chou, Timothy and Shaked, Tzook and Vontimitta, Varun and Ajayi, Victoria and Montanez, Victoria and Mohan, Vijai and Kumar, Vinay Satish and Mangla, Vishal and Albiero, Vítor and Ionescu, Vlad and Poenaru, Vlad and Mihailescu, Vlad Tiberiu and Ivanov, Vladimir and Li, Wei and Wang, Wenchen and Jiang, Wenwen and Bouaziz, Wes and Constable, Will and Tang, Xiaocheng and Wang, Xiaofang and Wu, Xiaojian and Wang, Xiaolan and Xia, Xide and Wu, Xilun and Gao, Xinbo and Chen, Yanjun and Hu, Ye and Jia, Ye and Qi, Ye and Li, Yenda and Zhang, Yilin and Zhang, Ying and Adi, Yossi and Nam, Youngjin and Yu and Wang and Hao, Yuchen and Qian, Yundi and He, Yuzi and Rait, Zach and DeVito, Zachary and Rosnbrick, Zef and Wen, Zhaoduo and Yang, Zhenyu and Zhao, Zhiwei},
	month = aug,
	year = {2024},}

@article{shahriar_putting_2024,
	title = {Putting {GPT}-4o to the {Sword}: {A} {Comprehensive} {Evaluation} of {Language}, {Vision}, {Speech}, and {Multimodal} {Proficiency}},
	volume = {14},	issn = {2076-3417},	number = {17},	journal = {Applied Sciences},
	author = {Shahriar, Sakib and Lund, Brady D. and Mannuru, Nishith Reddy and Arshad, Muhammad Arbab and Hayawi, Kadhim and Bevara, Ravi Varma Kumar and Mannuru, Aashrith and Batool, Laiba},
	month = jan,
	year = {2024},	pages = {7782},
}

@article{gu_mentalblend_2024,
	title = {{MentalBlend}: {Enhancing} {Online} {Mental} {Health} {Support} through the {Integration} of {LLMs} with {Psychological} {Counseling} {Theories}},
	volume = {46},	number = {0},	journal = {Proceedings of the Annual Meeting of the Cognitive Science Society},
	author = {Gu, Ziyin and Zhu, Qingmeng},
	year = {2024},
}

@misc{zhang_systematic_2024,
	title = {A {Systematic} {Survey} of {Text} {Summarization}: {From} {Statistical} {Methods} to {Large} {Language} {Models}},	publisher = {arXiv},
	author = {Zhang, Haopeng and Yu, Philip S. and Zhang, Jiawei},
	month = jun,
	year = {2024},}

@inproceedings{kocbek_generating_2022,	title = {Generating {Extremely} {Short} {Summaries} from the {Scientific} {Literature} to {Support} {Decisions} in {Primary} {Healthcare}: {A} {Human} {Evaluation} {Study}},	booktitle = {Artificial {Intelligence} in {Medicine}},
	publisher = {Springer International Publishing},
	author = {Kocbek, Primoz and Gosak, Lucija and Musović, Kasandra and Stiglic, Gregor},
	editor = {Michalowski, Martin and Abidi, Syed Sibte Raza and Abidi, Samina},
	year = {2022},	pages = {373--382},
}

@article{rao_single-document_2024,
	title = {Single-{Document} {Abstractive} {Text} {Summarization}: {A} {Systematic} {Literature} {Review}},
	volume = {57},
	issn = {0360-0300},	number = {3},	journal = {ACM Comput. Surv.},
	author = {Rao, Abishek and Aithal, Shivani and Singh, Sanjay},
	month = nov,
	year = {2024},
	pages = {60:1--60:37},
}

@inproceedings{zhang_this_2019,	title = {This {Email} {Could} {Save} {Your} {Life}: {Introducing} the {Task} of {Email} {Subject} {Line} {Generation}},	booktitle = {Proceedings of the 57th {Annual} {Meeting} of the {Association} for {Computational} {Linguistics}},
	publisher = {Association for Computational Linguistics},
	author = {Zhang, Rui and Tetreault, Joel},
	editor = {Korhonen, Anna and Traum, David and Màrquez, Lluís},
	month = jul,
	year = {2019},
	pages = {446--456},
}

@misc{jiang_mixtral_2024,
	title = {Mixtral of {Experts}},	publisher = {arXiv},
	author = {Jiang, Albert Q. and Sablayrolles, Alexandre and Roux, Antoine and Mensch, Arthur and Savary, Blanche and Bamford, Chris and Chaplot, Devendra Singh and Casas, Diego de las and Hanna, Emma Bou and Bressand, Florian and Lengyel, Gianna and Bour, Guillaume and Lample, Guillaume and Lavaud, Lélio Renard and Saulnier, Lucile and Lachaux, Marie-Anne and Stock, Pierre and Subramanian, Sandeep and Yang, Sophia and Antoniak, Szymon and Scao, Teven Le and Gervet, Théophile and Lavril, Thibaut and Wang, Thomas and Lacroix, Timothée and Sayed, William El},
	year = {2024},}

@misc{yin_novel_2024,
	title = {A {Novel} {LLM}-based {Two}-stage {Summarization} {Approach} for {Long} {Dialogues}},	publisher = {arXiv},
	author = {Yin, Yuan-Jhe and Chen, Bo-Yu and Chen, Berlin},
	month = oct,
	year = {2024},}

@article{zhang_benchmarking_2024,
	title = {Benchmarking {Large} {Language} {Models} for {News} {Summarization}},
	volume = {12},
	issn = {2307-387X},	journal = {Transactions of the Association for Computational Linguistics},
	author = {Zhang, Tianyi and Ladhak, Faisal and Durmus, Esin and Liang, Percy and McKeown, Kathleen and Hashimoto, Tatsunori B.},
	month = jan,
	year = {2024},
	pages = {39--57},
}

@inproceedings{lin_rouge_2004,	title = {{ROUGE}: {A} {Package} for {Automatic} {Evaluation} of {Summaries}},	booktitle = {Text {Summarization} {Branches} {Out}},
	publisher = {Association for Computational Linguistics},
	author = {Lin, Chin-Yew},
	month = jul,
	year = {2004},	pages = {74--81},
}

@inproceedings{zhang_bertscore_2019,
	title = {{BERTScore}: {Evaluating} {Text} {Generation} with {BERT}},	booktitle = {International {Conference} on {Learning} {Representations}},
	author = {Zhang, Tianyi and Kishore, Varsha and Wu, Felix and Weinberger, Kilian Q. and Artzi, Yoav},
	month = sep,
	year = {2019},}

@article{khan_comparison_2024,
	title = {A comparison of the diagnostic ability of large language models in challenging clinical cases},
	volume = {7},
	issn = {2624-8212},	journal = {Frontiers in Artificial Intelligence},
	author = {Khan, Maria Palwasha and O’Sullivan, Eoin Daniel},
	month = aug,
	year = {2024},	pages = {1379297},
}

@inproceedings{mirzakhmedova_are_2024,	title = {Are {Large} {Language} {Models} {Reliable} {Argument} {Quality} {Annotators}?},	booktitle = {Robust {Argumentation} {Machines}},
	publisher = {Springer Nature Switzerland},
	author = {Mirzakhmedova, Nailia and Gohsen, Marcel and Chang, Chia Hao and Stein, Benno},
	editor = {Cimiano, Philipp and Frank, Anette and Kohlhase, Michael and Stein, Benno},
	year = {2024},	pages = {129--146},
}

@article{hackl_is_2023,
	title = {Is {GPT}-4 a reliable rater? {Evaluating} consistency in {GPT}-4's text ratings},
	volume = {8},
	issn = {2504-284X},	journal = {Frontiers in Education},
	author = {Hackl, Veronika and Müller, Alexandra Elena and Granitzer, Michael and Sailer, Maximilian},
	month = dec,
	year = {2023},}

@book{krippendorff_content_2018,
	title = {Content {Analysis}: {An} {Introduction} to {Its} {Methodology}},	publisher = {SAGE Publications},
	author = {Krippendorff, Klaus},
	month = may,
	year = {2018},}

@inproceedings{papineni_bleu_2002,	title = {Bleu: a {Method} for {Automatic} {Evaluation} of {Machine} {Translation}},	booktitle = {Proceedings of the 40th {Annual} {Meeting} of the {Association} for {Computational} {Linguistics}},
	publisher = {Association for Computational Linguistics},
	author = {Papineni, Kishore and Roukos, Salim and Ward, Todd and Zhu, Wei-Jing},
	editor = {Isabelle, Pierre and Charniak, Eugene and Lin, Dekang},
	month = jul,
	year = {2002},	pages = {311--318},
}

@article{tam_framework_2024,
	title = {A framework for human evaluation of large language models in healthcare derived from literature review},
	volume = {7},	issn = {2398-6352},	number = {1},	journal = {npj Digital Medicine},
	author = {Tam, Thomas Yu Chow and Sivarajkumar, Sonish and Kapoor, Sumit and Stolyar, Alisa V. and Polanska, Katelyn and McCarthy, Karleigh R. and Osterhoudt, Hunter and Wu, Xizhi and Visweswaran, Shyam and Fu, Sunyang and Mathur, Piyush and Cacciamani, Giovanni E. and Sun, Cong and Peng, Yifan and Wang, Yanshan},
	month = sep,
	year = {2024},	pages = {1--20},
}

\end{document}